\documentstyle[maplems2]{article}
\begin{document}
\pagestyle{empty}

\begin{center}

\subsection*{Second-Order Black Hole Perturbations: A Computer Algebra
Approach \\
I - The Schwarzschild Spacetime}


\bf{George Davies}\\

\small{Department of Physics and Astronomy,}\\
\small{Queen's University, Kingston, Ontario K7L 3N6, Canada}\\
\small{davies@astro.queensu.ca} \\

\medskip

\today

\subsection*{Abstract}
\end{center}

This article outlines our derivation of the second order perturbations to a
Schwarzschild black hole, highlighting our use of, and necessary reliance on,
computer algebra. The particular perturbation scenario that is presented here
is the case of the linear quadrapole seeding the second order quadrapole. This
problem amounts to finding the second order Zerilli wave equation, and in
particular the effective source term due to the linear quadrapole.  With one
minor exception, our calculations confirm the earlier findings of 
Gleiser, $ et.$ $al.$ \cite{GL}. On route to these results we also illustrate
that, with the aid of computer algebra, the linear Schwarzschild problem can be
solved in a very direct manner ($i.e.$, without resorting to the usual function
transformations), and it is this ``direct method'' that drives the higher order
perturbation analysis.
The calculations were performed using the GRTensorII computer
algebra package, running on the Maple V platform, along with several 
new Maple routines that we have written specifically for these types of
problems. Although we have chosen to consider only the 
``quadrapole-quadrapole'' calculation in this article, the GRTensor
environment, with the inclusion of these new routines, would allow this
analysis to be repeated for a far more general problem.
These routines, along with Maple worksheets that reproduce our
calculations, are publicly available at the GRTensor website:
www.astro.queensu.ca/\verb+~+grtensor . The interested reader is invited to 
download and use them to reproduce our results and experiment.

\bigskip

\subsection*{Introduction and Overview}

The use of perturbative methods within general relativity to investigate the stability of black holes has, at the linear level, a long and 
successful history. The need for such analysis is obvious. While these objects are very mathematically rich and interesting, we need to 
know if they are indeed physical objects. If it were found that black holes were unstable to small perturbations we certainly would not 
be justified in expecting them to be the ultimate end state of runaway gravitational collapse. Of course, the linear analysis has shown 
that black holes are in fact stable within the context of first order perturbation theory. Regge and Wheeler \cite{RW} were the first to 
successfully perform a linear black hole analysis in the late 1950's with their ground-breaking study of the (spherically symmetric) 
Schwarzschild metric. Their results were later clarified and refined by Zerilli \cite{Z} in 1970, and it is this work that has become the 
standard for the linear analysis of the Schwarzschild problem. It is important to note that all of these calculations were carried out 
using the ``intuitive'' metric perturbation (MP) method, $i.e.$, by analyzing
perturbations of the metric tensor itself. This procedure 
will be described in greater detail below, but for now we simply note that this stands in stark contrast to the only successful linear 
analysis of the (rotating, axis symmetric) Kerr black hole. It is now not difficult to show by direct calculation that when one linearizes 
the Einstein equations about the Kerr vacuum solution the ``regular'' angular modes, $i.e.$, spherical harmonics, do not decouple. This 
is nothing less than a disaster for the metric perturbation analysis. Here one can no longer uniquely project out the contributions from 
each of the infinite number of multipoles required to construct a general perturbation.

The successful analysis of the linearized Kerr problem, first performed by Teukolsky \cite{T} in the early 1970's, avoids this problem of 
angular mode coupling entirely by recasting the problem in terms of the Neuman-Penrose (NP) formalism. (The NP formalism is an 
example of the more general tetrad approach to general relativity, that is distinguished by the fact that all of the tetrad vectors are 
chosen to be null. See for example, Chandrasekhar [5].) Much of the power that the NP approach brings to linear perturbation 
problems is a direct consequence of its formulation of the Einstein equations. Here six of the Einstein equations are linear(!) at the 
outset. An immediate result of this rather remarkable fact is that any quantities that are zero in the background solution automatically 
appear as linearized perturbations. The essence of this approach then, is that one does not have to linearize the Einstein equations at all. 
It should be noted that performing a ``useful'' linearization of the Einstein
equations about a given metric, even with the aid of modern computer algebra
packages, 
can be a non-trivial task. While the NP and MP methods can in principle be related to each other via the perturbed tetrad vectors, 
which must combine to form the perturbed metric functions, the two approaches can appear to have surprisingly little in common. We 
shall return to this point in our concluding remarks. As a final comment on the linearized Kerr analysis we note that while the NP 
scheme does make the problem tractable, the analysis remains rather complex, and is, in the words of Chandrasekhar \cite{Chandra}, 
``...prolixious in its complexity.''.

\pagestyle{myheadings}

The ultimate goal of this article is to present our calculations of second order perturbations to the Schwarzschild metric, with an 
emphasis on the computer algebra methods, packages, and techniques that we employ. To properly set the stage for this end result we 
must first outline the details of the linear perturbation theory and its connection to the second, and higher, order perturbations. This 
material appears in the following section. The next section will then give an example of the linear theory at work. Here we will follow 
through the linear Schwarzschild calculation and see that with the aid of modern computer algebra this analysis can be carried out in a 
thoroughly transparent fashion. Here we will not derive the usual Zerilli 
results, but rather illustrate, and find the prescription for, the decoupling 
of the linear perturbation modes.
With this material covered we will then move on to illustrate the second order
calculations. The main goal of this section will be to find the form of the 
effective potential in the Zerilli wave equation.
This is where we will see the absolute necessity of employing computer
algebraic methods to perform these types of calculations. We will also 
see that with the addition of a few new simplification routines the analysis
can be made much simpler. The final section will, of course, 
summarize our results and attempt to view them in the broader context of more
general problems. The current direction of our ongoing research will also be
discussed.

\subsection*{Perturbation Theory: A Short Review}

At the outset it is important that we clearly define what we mean by a perturbation, and its relative order. Equally important is that we 
come to fully understand the linear problem. Indeed, once the linear problem has been completely solved, we will see that the second 
order problem is, in principle, ``just'' a matter of computational volume. In what follows we will limit our discussion to the MP method, 
and give only a brief description of the methodology. (A discussion of some of the subtler points of the associated questions of gauge can be 
found in [1].) We start by writing the metric as a known, background, vacuum solution with the addition of higher order terms that 
mimic an expansion in a ``small'' perturbation parameter, $\epsilon$,
$$
g_{\alpha\beta}=g_{\alpha\beta}^{(0)}+\epsilon g_{\alpha\beta}^{(1)}+ \epsilon^2 g_{\alpha\beta}^{(2)}+ \dots~,
$$
and substitute this ``expansion'' into the vacuum Einstein equations,
$$
R_{\alpha\beta} = 0~.
$$
To produce an ``$n^{th}$'' order perturbation we simply truncate all terms with powers higher than $O(\epsilon^n)$ in the resulting 
equations; so that a first order perturbation involves keeping only up to  $O(\epsilon)$ terms.  This, along with our restriction to 
vacuum background solutions, allows us to express the linearly perturbed
Einstein equations as a homogeneous linear operator, that depends on 
the background metric, acting on the linear perturbations\footnote{This is really just the statement that the linearly perturbed Einstein 
equations reduce to, $R_{\alpha\beta}^{(1)}=0$, in vacuum.}: 
$$
\epsilon{\cal L}_{g_{\alpha\beta}^{(0)}}\left( g_{\alpha\beta}^{(1)}\right)=0 ~.
$$
The next step in the MP process is to perform the ``usual'' mode expansion of the perturbation functions,
$$
g^{(1)} = \sum_{i=0}^{\infty} h_{i}(r,t) u_{i}(\theta,\phi)~,
$$
where $g^{(1)}$ represents all of the perturbed metric functions, and the $u_{i}(\theta,\phi)$ are typically spherical harmonics or 
Legendre functions. If all goes well, when these expansions are put through the linear operator the angular modes will decouple. The 
mathematical statement of this is, 
$$
\left\langle f_1(u_j) \left| {\cal L} \left(\sum_{i=0}^{\infty} h_{i}u_{i}\right)\right. \right\rangle = f_2(h_j)~,
$$
where $f_1$ and $f_2$ are some functions of their arguments and their derivatives\footnote{Recall, for instance, that for Legendre functions, 
$\langle u_i|u_j\rangle=\int_0^\pi d\theta sin(\theta) u_i u_j \propto \delta_{ij}$.}. Essentially, we need a prescription for picking out the 
contributions that a particular $h_j$ makes to the linearized equations. As one might guess, such a unique decomposition is not always 
possible.

The nature of any decoupling will of course depend on the symmetries of the background metric, which determines the functional form 
of ${\cal L}$. Thus, one might also correctly guess that the spherical symmetry of the background Schwarzschild metric will allow for 
the separation of spherical harmonic modes, while the background Kerr metric, which
obeys the far more restrictive axial symmetry, will not. As a preview of the next section we note that in the linear Schwarzschild 
analysis, three of the perturbed Ricci components take the form,
$$
{\cal L}_{Schw}\left(\sum_{i=0}^{\infty} h_{i}u_{i}\right) = \sum_{i=0}^{\infty}f_2(h_i)~u_i = 0~,
$$ 
so that a simple projection of $\langle u_j |$ will yield $f_2(h_j)=0$, which is one of the desired PDEs governing the metric 
perturbations. For the simpler case of decoupling modes then, the angular functions are known and the problem is then to determine the 
differential equations that the $h_i$ obey. Whereas, in the more general case the angular eigenfunctions must themselves be determined 
as part of the solution. For the case of the linearized Kerr metric, for instance, these angular functions can be shown to form a general 
Sturm-Liouville system, but they must be determined numerically, see $e.g.$, \cite{T}. 

Assuming that we have completely solved the linearized problem, $i.e.$, once the angular eigenfunctions and the PDEs relating the 
$h_i$, are known, we can begin to quantitatively examine the second order perturbations. Consider truncating the main perturbation 
equation at $O(\epsilon^3)$.
It is not hard to convince oneself that the $O(\epsilon^2)$ terms occuring in
the perturbed Einstein equations now yield the $inhomogeneous$ system,
$$
\epsilon^2{\cal L}_{g_{\alpha\beta}^{(0)}}\left( g_{\alpha\beta}^{(2)}\right)={\cal S}\left(\epsilon^2 
\left(g_{\alpha\beta}^{(1)}\right)^2\right)~,
$$ 
where ${\cal L}$ is exactly the same linear operator found in the first order analysis, and ${\cal S}$ is just the collection of 
quadratically occuring linear perturbations (assumed known from the linear
analysis) that are produced by keeping all of the $O(\epsilon^2)$ terms 
that appear in the perturbed Einstein equations. The physical significance of
this inhomogeneous relationship, and the need 
for a clear understanding of the linear analysis, is now apparent. The
quadratic, first order perturbation, terms act as an effective source term for
the second 
order perturbations, allowing us to quantify the back reaction of the linear perturbations on the system. The determination of the PDEs 
for the second order perturbations now follows exactly the same procedure as in the linear analysis, but here the projections of the 
angular eigenfunctions will pick out components of the source terms from ${\cal S}$. Note that without a clear analysis of the linear 
problem one would not know how to calculate what the projected components of the source terms are. While this may seem obvious, 
much of the linear Schwarzschild problem can be solved by simply $assuming$ that the angular modes decouple, and setting various 
coefficients of $\cos^{n}(\theta)$ equal to zero. If one chooses this, albeit simpler, solution technique for the linear problem the second 
order analysis is doomed.

This completes our general discussion of MP methods, and we now move on to 
examine the particular case of perturbations to the Schwarzschild spacetime.
The next section will demonstrate, with the aid of computer algebra, the 
decoupling of the linear Schwarzschild modes. As we shall see, the use of computer algebra allows the linear analysis to be examined in 
a thoroughly transparent manner. With the computer doing all of the work, we do not have to resort to invoking the standard function
transformations, which mix the original perturbation functions, to simplify the calculations. The final result of the analysis 
presented in the next section will be a simple procedure for extracting the
decoupled perturbation modes, of arbitrary order, for a Schwarzschild black
hole.

\subsection*{The Linear Schwarzschild Analysis}

The bulk of this section is the input/output from a GRTensorII session run on
the Maple V platform. (The Maple worksheet itself will be made available for
download from the GRTensor website.)
While reading this section one 
should bear in mind what our objective is. Rather than actually deriving the standard linear Schwarzschild results, $i.e.$ the Zerilli 
wave equation \cite{Z}, we simply want to show that the angular modes of the linearized Schwarzschild metric will decouple. In so 
doing we will develop a procedure to consistently extract a particular multipole from the perturbed Ricci tensor. Note that in what 
follows we consider only a single ``$i^{th}$'' term from the general perturbation expansion presented in the last section, with its index 
suppressed. This is done for simplicity and clarity, and with no loss of generality due to the linearity of ${\cal L}$.  (The concerned 
reader can mentally place a summed index on all of the perturbation terms.) 


Our general approach to this problem begins by constructing the standard
form of the covariant, linearly perturbed, Schwarzschild metric,
$$
g_{\alpha\beta} \simeq g_{\alpha\beta}^{(0)}+ \epsilon g_{\alpha\beta}^{(1)}~,
$$
where,
$$
g_{\alpha\beta}^{(0)}=
diag\left(g_{rr},g_{\theta\theta},g_{\phi\phi},g_{tt}\right)=
diag\left((1-2M/r)^{-1},r^2,r^2\sin^2(\theta),-(1-2M/r)\right)~.
$$
For a discussion of the 
particular form of the perturbed metric that we use, which will be seen
shortly, see 
$e.g.$ [1].
The next step in this process is to
calculate the {\em exact} contravariant metric tensor, $g^{\alpha\beta}$, that
is
associated
with the linearly perturbed covariant metric, $g_{\alpha\beta}$, and then to 
linearize it
w.r.t. $\epsilon$. The resulting object is the linearly perturbed
contravariant metric tensor,
$$
g^{\alpha\beta}\simeq g_{(0)}^{\alpha\beta}+\epsilon g_{(1)}^{\alpha\beta} =
g_{(0)}^{\alpha\beta}+\left.\left(\frac{\partial g^{\alpha\beta}}
{\partial \epsilon}\right|_{\epsilon=0}\right)\epsilon ~.
$$
These linearized co/contra-variant forms of the metric tensor are the
fundamental tools of the linear theory: With them one can calculate the linear
perturbation to any tensor
quantity by simply following through the normal, unperturbed, calculation
while dropping all of the non-linear $\epsilon$ terms that appear.
While this
procedure can certainly produce calculations that would be intractable
``by hand'', it is a
very simple matter to instruct the Maple computer engine to follow this
algorithmic procedure. To arrive at the linearized Einstein vacuum equations
then, we
simply instruct GRTensor to calculate the Ricci tensor from the linearly
perturbed metric, while truncating the non-linear $\epsilon$ terms.
$$
R_{\alpha\beta}\left(g_{\gamma\delta}^{(0)}+ \epsilon g_{\gamma\delta}^{(1)},
g_{(0)}^{\gamma\delta} +\epsilon g_{(1)}^{\gamma\delta}\right)=
R_{\alpha\beta}^{(0)}+ \epsilon R_{\alpha\beta}^{(1)}+O(\epsilon^2)=0~.
$$
Of course, $R_{\alpha\beta}^{(0)}=0$, for the Schwarzschild spacetime, so that
truncation of the non-linear $\epsilon$ term yields the linearized Einstein
equations, $R_{\alpha\beta}^{(1)}=0$.

To begin the Maple session we load the GRTensorII libraries, our new simplification/perturbation routines\footnote{These routines are 
contained in the ``myutils.mpl'' file and are detailed in Appendix A.}, and the linearly perturbed metric.

\bigskip

\begin{mapleinput}
restart:
\end{mapleinput}

\begin{mapleinput}
readlib(grii):
\end{mapleinput}
\begin{mapleinput}
grtensor():
\end{mapleinput}
\begin{maplelatex}
\[
{\it GRTensorII\:Version\:1.64\:(R3)}
\]
\end{maplelatex}
\begin{maplelatex}
\[
{\it 4\:November\:1997}
\]
\end{maplelatex}
\begin{maplelatex}
\[
{\it Developed\:by\:Peter\:Musgrave,\:Denis\:Pollney\:and\:Kayll
\:Lake}
\]
\end{maplelatex}
\begin{maplelatex}
\[
{\it Copyright\:1994-1997\:by\:the\:authors.}
\]
\end{maplelatex}
\begin{maplelatex}
\[
{\it Latest\:version\:available\:from:\:http://astro.queensu.ca/
\char'176grtensor/}
\]
\end{maplelatex}
\begin{maplelatex}
\[
{\it To\:initiate\:help\:type\:?grtensor}
\]
\end{maplelatex}

\begin{mapleinput}
mine():
\end{mapleinput}
\begin{mapleinput}
read `myutils.mpl`;
\end{mapleinput}

\begin{mapleinput}
qload(lpschw):
\end{mapleinput}
\begin{maplettyout}
Calculating ds for lpschw ... Done. (0.000000 sec.) 

\end{maplettyout}
\begin{maplelatex}
\[
{\it Default\:spacetime}={\it lpschw}
\]
\end{maplelatex}
\begin{maplelatex}
\[
{\it For\:the\:lpschw\:spacetime:}
\]
\end{maplelatex}
\begin{maplelatex}
\[
{\it Coordinates}
\]
\end{maplelatex}
\begin{maplelatex}
\[
{\rm x}(\,{\it up}\,)
\]
\end{maplelatex}
\begin{maplelatex}
\[
{\it x\:}^{{a}}=[\,{r}\,{ \theta}\,{ \phi}\,{t}\,]
\]
\end{maplelatex}
\begin{maplelatex}
\[
{\it Line\:element}
\]
\end{maplelatex}
\begin{maplelatex}
\begin{eqnarray*}
\lefteqn{{\it \:ds}^{2}={\displaystyle \frac { \left( \! \,1 + { 
\varepsilon}\,{{H}_{2}}(\,{r}, {t}\,)\,{\rm u}(\,{ \theta}\,)\,
 \!  \right) \,{\it \:d}\,{r}^{{\it 2\:}}}{1 - 2\,{\displaystyle 
\frac {{m}}{{r}}}}} + 2\,{ \varepsilon}\,{{H}_{1}}(\,{r}, {t}\,)
\,{\rm u}(\,{ \theta}\,)\,{\it \:d}\,{r}^{{\:}}\,{\it d\:}\,{t}^{
{\:}}} \\
 & & \mbox{} + {r}^{2}\,(\,1 + { \varepsilon}\,{\rm K}(\,{r}, {t}
\,)\,{\rm u}(\,{ \theta}\,)\,)\,{\it \:d}\,{ \theta}^{{\it 2\:}}
 + {r}^{2}\,{\rm sin}(\,{ \theta}\,)^{2}\,(\,1 + { \varepsilon}\,
{\rm K}(\,{r}, {t}\,)\,{\rm u}(\,{ \theta}\,)\,)\,{\it \:d}\,{ 
\phi}^{{\it 2\:}} \\
 & & \mbox{} -  \left( \! \,1 - 2\,{\displaystyle \frac {{m}}{{r}
}}\, \!  \right) \, \left( \! \,1 - { \varepsilon}\,{{H}_{0}}(\,{
r}, {t}\,)\,{\rm u}(\,{ \theta}\,)\, \!  \right) \,{\it \:d}\,{t}
^{{\it 2\:}}
\end{eqnarray*}
\end{maplelatex}
\begin{maplelatex}
\[
{\it Constraints}= \left[ \! \,{ \varepsilon}^{2}=0, { 
\varepsilon}^{3}=0, { \varepsilon}^{4}=0, { \varepsilon}^{5}=0, {
 \varepsilon}^{6}=0, { \varepsilon}^{7}=0, { \varepsilon}^{8}=0, 
{ \varepsilon}^{9}=0, { \varepsilon}^{10}=0\, \!  \right] 
\]
\end{maplelatex}

\noindent These constraints will allow us to truncate the non-linear perturbation terms during calculation, which greatly reduces both 
the CPU and memory requirements. Note that the function $u(\theta)$ appearing in the metric represents an arbitrary Legendre 
function. (Although, it will actually remain an arbitrary function of $\theta$ until we eliminate $u^{\prime\prime}$ in terms of 
$u^{\prime}$ and $u$, using the Legendre equation, later in the calculation.)

Our analysis begins with the calculation of the exact $g^{\alpha\beta}$ corresponding to the input metric, which we must then 
linearize in $\epsilon$. The linearization is accomplished using the
{\bf linpert()} routine, the details of which can be found in Appendix A.
We are then left with, 
$g_{\alpha\beta}= g^{(0)}_{\alpha\beta}+\epsilon g^{(1)}_{\alpha\beta}$, and,
$g^{\alpha\beta}= {g_{(0)}}^{\alpha\beta}+\epsilon {g_{(1)}}^{\alpha\beta}$,
as the co/contra-variant components of the metric tensor\footnote{Direct 
calculation shows that, $g^{\alpha\gamma}g_{\gamma\beta} = \delta^\alpha_\beta + O(\epsilon^2)$, as required for consistency at 
$O(\epsilon)$.}.

\bigskip

\begin{mapleinput}
grcalc(g(up,up)); 
\end{mapleinput}
\begin{maplettyout}
Calculating detg for lpschw ... Done. (0.017000 sec.) 
Calculating g(up,up) for lpschw ... Done. (0.067000 sec.) 

\end{maplettyout}
\begin{maplelatex}
\[
{\it CPU\:Time\:}=.067
\]
\end{maplelatex}

\begin{mapleinput}
gralter(g(up,up),linpert,simplify,factor); 
\end{mapleinput}
\begin{maplettyout}
Component simplification of a GRTensorII object:
 
Applying routine linpert to object g(up,up)
Applying routine simplify to object g(up,up)
Applying routine factor to object g(up,up)

\end{maplettyout}
\begin{maplelatex}
\[
{\it CPU\:Time\:}=.133
\]
\end{maplelatex}
\begin{mapleinput}
grdisplay(g(up,up)); 
\end{mapleinput}
\begin{maplelatex}
\[
{\it For\:the\:lpschw\:spacetime:}
\]
\end{maplelatex}
\begin{maplelatex}
\[
{\it Contravariant\:metric\:tensor}
\]
\end{maplelatex}
\begin{maplelatex}
\[
{\rm g}(\,{\it up}, {\it up}\,)
\]
\end{maplelatex}
\begin{maplelatex}
\[
{\it g\:}^{{a}}\,{}^{{b}}= \left[ 
{\begin{array}{c}
 - \,{\displaystyle \frac { \left( \! \, - 1 + { \varepsilon}\,{{
H}_{2}}(\,{r}, {t}\,)\,{\rm u}(\,{ \theta}\,)\, \!  \right) \,(\,
{r} - 2\,{m}\,)}{{r}}}\,, 0\,, 0\,, { \varepsilon}\,{{H}_{1}}(\,{
r}, {t}\,)\,{\rm u}(\,{ \theta}\,) \\ [2ex]
0\,, \, - \,{\displaystyle \frac { - 1 + { \varepsilon}\,{\rm K}(
\,{r}, {t}\,)\,{\rm u}(\,{ \theta}\,)}{{r}^{2}}}\,, \,0\,, \,0 \\
 [2ex]
0\,, \,0\,, \,{\displaystyle \frac { - 1 + { \varepsilon}\,{\rm K
}(\,{r}, {t}\,)\,{\rm u}(\,{ \theta}\,)}{{r}^{2}\,(\,{\rm cos}(\,
{ \theta}\,) - 1\,)\,(\,{\rm cos}(\,{ \theta}\,) + 1\,)}}\,, \,0
 \\ [2ex]
{ \varepsilon}\,{{H}_{1}}(\,{r}, {t}\,)\,{\rm u}(\,{ \theta}\,)\,
, \,0\,, \,0\,, \, - \,{\displaystyle \frac {{r}\, \left( \! \,1
 + { \varepsilon}\,{{H}_{0}}(\,{r}, {t}\,)\,{\rm u}(\,{ \theta}\,
)\, \!  \right) }{{r} - 2\,{m}}}
\end{array}}
 \right] 
\]
\end{maplelatex}

With the linearized co/contra-variant forms of the metric tensor in hand, it is a simple matter to calculate the perturbed Ricci tensor, or 
any other first order (tensor) quantities of interest for that matter.
We now calculate the Ricci tensor, which amounts to calculating the perturbed Einstein tensor, 
applying the constraints as we go to kill off the higher order terms.

\bigskip

\begin{mapleinput}
grcalcalter(R(dn,dn),13); 
\end{mapleinput}
\begin{maplettyout}
Simplification will be applied during calculation.
 
Applying routine Apply constraints repeatedly to object g(dn,dn,pdn)
Calculating g(dn,dn,pdn) for lpschw ... Done. (0.033000 sec.) 
Applying routine Apply constraints repeatedly to object Chr(dn,dn,dn)
Calculating Chr(dn,dn,dn) for lpschw ... Done. (0.050000 sec.) 
Applying routine Apply constraints repeatedly to object Chr(dn,dn,up)
Calculating Chr(dn,dn,up) for lpschw ... Done. (0.117000 sec.) 
Applying routine Apply constraints repeatedly to object R(dn,dn)
Calculating R(dn,dn) for lpschw ... Done. (0.733000 sec.) 

\end{maplettyout}
\begin{maplelatex}
\[
{\it CPU\:Time\:}=.933
\]
\end{maplelatex}

\noindent Up to this point the angular function $u(\theta)$ has been completely arbitrary, except that one would like it to be
a member of a complete set. We now fix $u$ as a Legendre function by making the following substitution in $R_{\alpha\beta}$: 
$u^{\prime\prime} =- \cot(\theta)u^\prime - n(n+1)u = 0$. (For simplicity we set $j=n(n+1)$.)

\bigskip

\begin{mapleinput}
grmap(R(dn,dn),subs,diff(u(theta),theta$2)=
-cos(theta)/sin(theta)*diff(u(theta),theta)-j*u(theta),`x`); 
\end{mapleinput}
\begin{maplettyout}
Applying routine subs to R(dn,dn)

\end{maplettyout}
\noindent We can now examine the seven non-trivial components of the perturbed Ricci tensor: $R_{r \theta}^{(1)}$, $R_{r t}^{(1)}$, 
$R_{\theta t}^{(1)}$, and the four diagonal components. The
$R_{r \theta}^{(1)}$ component is\footnote{Although 
$R^{(0)}_{\alpha\beta}=0$, we will still take the time to collect only the linear $\epsilon$ terms as a matter of good practice.}:

\bigskip

\begin{mapleinput}
b1:=hcollect(expand(coeff(collect(grcomponent(R(dn,dn),[r,theta]),epsilon),epsilon,1)),
ufuncs,lperts);
\end{mapleinput}
\begin{maplelatex}
\begin{eqnarray*}
\lefteqn{{\it b1} :=  \left( {\vrule 
height1.22em width0em depth1.22em} \right. \! \! {\displaystyle 
\frac {1}{2}}\,{\displaystyle \frac {(\, - {m} + {r}\,)\,{{H}_{2}
}(\,{r}, {t}\,)}{(\,{r} - 2\,{m}\,)\,{r}}} - {\displaystyle 
\frac {1}{2}}\,{\displaystyle \frac {(\, - 3\,{m} + {r}\,)\,{{H}
_{0}}(\,{r}, {t}\,)}{(\,{r} - 2\,{m}\,)\,{r}}} + {\displaystyle 
\frac {1}{2}}\, \left( \! \,{\frac {{ \partial}}{{ \partial}{r}}}
\,{{H}_{0}}(\,{r}, {t}\,)\, \!  \right)  - {\displaystyle \frac {
1}{2}}\, \left( \! \,{\frac {{ \partial}}{{ \partial}{r}}}\,{\rm 
K}(\,{r}, {t}\,)\, \!  \right) } \\
 & & \mbox{} - {\displaystyle \frac {1}{2}}\,{\displaystyle 
\frac {{r}\, \left( \! \,{\frac {{ \partial}}{{ \partial}{t}}}\,{
{H}_{1}}(\,{r}, {t}\,)\, \!  \right) }{{r} - 2\,{m}}} \! 
\! \left. {\vrule height1.22em width0em depth1.22em} \right) 
 \left( \! \,{\frac {{ \partial}}{{ \partial}{ \theta}}}\,{\rm u}
(\,{ \theta}\,)\, \!  \right) \mbox{\hspace{222pt}}
\end{eqnarray*}
\end{maplelatex}

\noindent (The {\bf hcollect()} routine, described in Appendix A, is a hierarchical collection procedure, ``ufuncs'' is the set of $u$ and its 
derivatives, and ``lperts'' is the set of linear MP and their derivatives up to second order.) If we considered the full series perturbation 
mentioned above, we would have an infinite sum of these expressions, each with index $i$. Noting that the $u_i^\prime(\theta)$  form 
an orthogonal set however, we could uniquely select out any desired term in the imagined summation by taking the appropriate 
projection on the linearized expression (and any potential source term). Examining $R_{t \theta}^{(1)}$ we similarly find:

\bigskip

\begin{mapleinput}
c1:=hcollect(coeff(collect(grcomponent(R(dn,dn),[t,theta]),epsilon),epsilon,1),
ufuncs,lperts);
\end{mapleinput}
\begin{maplelatex}
\[
{\it c1} :=  \left( \! \,{\displaystyle \frac {{{H}_{1}}(\,{r}, {
t}\,)\,{m}}{{r}^{2}}} + {\displaystyle \frac {1}{2}}\,
{\displaystyle \frac {(\,{r} - 2\,{m}\,)\, \left( \! \,{\frac {{ 
\partial}}{{ \partial}{r}}}\,{{H}_{1}}(\,{r}, {t}\,)\, \! 
 \right) }{{r}}} - {\displaystyle \frac {1}{2}}\, \left( \! \,
{\frac {{ \partial}}{{ \partial}{t}}}\,{{H}_{2}}(\,{r}, {t}\,)\,
 \!  \right)  - {\displaystyle \frac {1}{2}}\, \left( \! \,
{\frac {{ \partial}}{{ \partial}{t}}}\,{\rm K}(\,{r}, {t}\,)\,
 \!  \right) \, \!  \right) \, \left( \! \,{\frac {{ \partial}}{{
 \partial}{ \theta}}}\,{\rm u}(\,{ \theta}\,)\, \!  \right) 
\]
\end{maplelatex}

\noindent Here again we simply need to project with $u_i^\prime(\theta)$ in order to extract the ``$i^{th}$'' term from any summed 
expression. We next find that $R_{r t}^{(1)}$, $R_{r r}^{(1)}$, 
and $R_{t t}^{(1)}$ can all be expressed as: $f(h_i) u_i(\theta)=0$. 
For these three components we need only project with the $u_i$ function itself.

\bigskip

\begin{mapleinput}
d1:=hcollect(expand(coeff(collect(grcomponent(R(dn,dn),[t,r]),epsilon),epsilon,1)),
ufuncs,lperts);
\end{mapleinput}
\begin{maplelatex}
\[
{\it d1} :=  \left( \! \,{\displaystyle \frac {1}{2}}\,
{\displaystyle \frac {{{H}_{1}}(\,{r}, {t}\,)\,{j}}{{r}^{2}}} - 
 \left( \! \,{\frac {{ \partial}^{2}}{{ \partial}{r}\,{ \partial}
{t}}}\,{\rm K}(\,{r}, {t}\,)\, \!  \right)  + {\displaystyle 
\frac {{\frac {{ \partial}}{{ \partial}{t}}}\,{{H}_{2}}(\,{r}, {t
}\,)}{{r}}} - {\displaystyle \frac {(\, - 3\,{m} + {r}\,)\,
 \left( \! \,{\frac {{ \partial}}{{ \partial}{t}}}\,{\rm K}(\,{r}
, {t}\,)\, \!  \right) }{(\,{r} - 2\,{m}\,)\,{r}}}\, \!  \right) 
\,{\rm u}(\,{ \theta}\,)
\]
\end{maplelatex}

\begin{mapleinput}
e1:=hcollect(expand(coeff(collect(grcomponent(R(dn,dn),[r,r]),epsilon),epsilon,1)),
ufuncs,lperts);
\end{mapleinput}
\begin{maplelatex}
\begin{eqnarray*}
\lefteqn{{\it e1} :=  \left( {\vrule 
height1.29em width0em depth1.29em} \right. \! \! {\displaystyle 
\frac {1}{2}}\,{\displaystyle \frac {{{H}_{2}}(\,{r}, {t}\,)\,{j}
}{{r}\,(\,{r} - 2\,{m}\,)}} + {\displaystyle \frac {1}{2}}\,
{\displaystyle \frac {{r}^{2}\, \left( \! \,{\frac {{ \partial}^{
2}}{{ \partial}{t}^{2}}}\,{{H}_{2}}(\,{r}, {t}\,)\, \!  \right) 
}{(\,{r} - 2\,{m}\,)^{2}}} - {\displaystyle \frac { \left( \! \,
{\frac {{ \partial}^{2}}{{ \partial}{r}\,{ \partial}{t}}}\,{{H}_{
1}}(\,{r}, {t}\,)\, \!  \right) \,{r}}{{r} - 2\,{m}}} + 
{\displaystyle \frac {1}{2}}\, \left( \! \,{\frac {{ \partial}^{2
}}{{ \partial}{r}^{2}}}\,{{H}_{0}}(\,{r}, {t}\,)\, \!  \right) }
 \\
 & & \mbox{} -  \left( \! \,{\frac {{ \partial}^{2}}{{ \partial}{
r}^{2}}}\,{\rm K}(\,{r}, {t}\,)\, \!  \right)  + {\displaystyle 
\frac {3}{2}}\,{\displaystyle \frac { \left( \! \,{\frac {{ 
\partial}}{{ \partial}{r}}}\,{{H}_{0}}(\,{r}, {t}\,)\, \! 
 \right) \,{m}}{{r}\,(\,{r} - 2\,{m}\,)}} + {\displaystyle 
\frac {1}{2}}\,{\displaystyle \frac {(\, - 3\,{m} + 2\,{r}\,)\,
 \left( \! \,{\frac {{ \partial}}{{ \partial}{r}}}\,{{H}_{2}}(\,{
r}, {t}\,)\, \!  \right) }{{r}\,(\,{r} - 2\,{m}\,)}} \\
 & & \mbox{} - {\displaystyle \frac {(\, - 3\,{m} + 2\,{r}\,)\,
 \left( \! \,{\frac {{ \partial}}{{ \partial}{r}}}\,{\rm K}(\,{r}
, {t}\,)\, \!  \right) }{{r}\,(\,{r} - 2\,{m}\,)}} - 
{\displaystyle \frac {{m}\, \left( \! \,{\frac {{ \partial}}{{ 
\partial}{t}}}\,{{H}_{1}}(\,{r}, {t}\,)\, \!  \right) }{(\,{r} - 
2\,{m}\,)^{2}}} \! \! \left. {\vrule 
height1.29em width0em depth1.29em} \right) {\rm u}(\,{ \theta}\,)
\mbox{\hspace{100pt}}
\end{eqnarray*}
\end{maplelatex}

\begin{mapleinput}
f1:=hcollect(expand(coeff(collect(grcomponent(R(dn,dn),[t,t]),epsilon),epsilon,1)),
ufuncs,lperts);
\end{mapleinput}
\begin{maplelatex}
\begin{eqnarray*}
\lefteqn{{\it f1} :=  \left( {\vrule 
height1.22em width0em depth1.22em} \right. \! \!  - \,
{\displaystyle \frac {1}{2}}\, \left( \! \,{\frac {{ \partial}^{2
}}{{ \partial}{t}^{2}}}\,{{H}_{2}}(\,{r}, {t}\,)\, \!  \right) 
 -  \left( \! \,{\frac {{ \partial}^{2}}{{ \partial}{t}^{2}}}\,
{\rm K}(\,{r}, {t}\,)\, \!  \right)  + {\displaystyle \frac {(\,{
r} - 2\,{m}\,)\, \left( \! \,{\frac {{ \partial}^{2}}{{ \partial}
{r}\,{ \partial}{t}}}\,{{H}_{1}}(\,{r}, {t}\,)\, \!  \right) }{{r
}}}} \\
 & & \mbox{} - {\displaystyle \frac {1}{2}}\,{\displaystyle 
\frac {(\,{r} - 2\,{m}\,)^{2}\, \left( \! \,{\frac {{ \partial}^{
2}}{{ \partial}{r}^{2}}}\,{{H}_{0}}(\,{r}, {t}\,)\, \!  \right) 
}{{r}^{2}}} + {\displaystyle \frac {1}{2}}\,{\displaystyle 
\frac {(\,{r} - 2\,{m}\,)\,{{H}_{0}}(\,{r}, {t}\,)\,{j}}{{r}^{3}
}} \\
 & & \mbox{} - {\displaystyle \frac {1}{2}}\,{\displaystyle 
\frac {(\,2\,{r} - {m}\,)\,(\,{r} - 2\,{m}\,)\, \left( \! \,
{\frac {{ \partial}}{{ \partial}{r}}}\,{{H}_{0}}(\,{r}, {t}\,)\,
 \!  \right) }{{r}^{3}}} - {\displaystyle \frac {1}{2}}\,
{\displaystyle \frac {{m}\,(\,{r} - 2\,{m}\,)\, \left( \! \,
{\frac {{ \partial}}{{ \partial}{r}}}\,{{H}_{2}}(\,{r}, {t}\,)\,
 \!  \right) }{{r}^{3}}} \\
 & & \mbox{} + {\displaystyle \frac {(\,{r} - 2\,{m}\,)\,{m}\,
 \left( \! \,{\frac {{ \partial}}{{ \partial}{r}}}\,{\rm K}(\,{r}
, {t}\,)\, \!  \right) }{{r}^{3}}} + {\displaystyle \frac {(\, - 
3\,{m} + 2\,{r}\,)\, \left( \! \,{\frac {{ \partial}}{{ \partial}
{t}}}\,{{H}_{1}}(\,{r}, {t}\,)\, \!  \right) }{{r}^{2}}} \! 
\! \left. {\vrule height1.22em width0em depth1.22em} \right) 
{\rm u}(\,{ \theta}\,)
\end{eqnarray*}
\end{maplelatex}

\noindent So far all of the perturbed Ricci components have had a very simple angular dependence in terms of either, $u(\theta)$, 
or its derivative. The last two non-trivial components of 
$R_{\alpha\beta}^{(1)}$ do not, however, exhibit this simplicity.

\bigskip

\begin{mapleinput}
a1:=hcollect(coeff(collect(grcomponent(R(dn,dn),[theta,theta]),epsilon),epsilon,1),
ufuncs,lperts);
\end{mapleinput}
\begin{maplelatex}
\begin{eqnarray*}
\lefteqn{{\it a1} :=  \left( {\vrule 
height1.22em width0em depth1.22em} \right. \! \! {\displaystyle 
\frac {1}{2}}\,(\,{j} + 2\,)\,{{H}_{2}}(\,{r}, {t}\,) + 
{\displaystyle \frac {1}{2}}\,(\, - 2 + {j}\,)\,{\rm K}(\,{r}, {t
}\,) + {\displaystyle \frac {1}{2}}\,{\displaystyle \frac {{r}^{3
}\, \left( \! \,{\frac {{ \partial}^{2}}{{ \partial}{t}^{2}}}\,
{\rm K}(\,{r}, {t}\,)\, \!  \right) }{{r} - 2\,{m}}}} \\
 & & \mbox{} - {\displaystyle \frac {1}{2}}\, \left( \! \,
{\frac {{ \partial}^{2}}{{ \partial}{r}^{2}}}\,{\rm K}(\,{r}, {t}
\,)\, \!  \right) \,(\,{r} - 2\,{m}\,)\,{r} - {\displaystyle 
\frac {1}{2}}\,{{H}_{0}}(\,{r}, {t}\,)\,{j} + {\displaystyle 
\frac {1}{2}}\, \left( \! \,{\frac {{ \partial}}{{ \partial}{r}}}
\,{{H}_{0}}(\,{r}, {t}\,)\, \!  \right) \,(\,{r} - 2\,{m}\,) \\
 & & \mbox{} + {\displaystyle \frac {1}{2}}\, \left( \! \,
{\frac {{ \partial}}{{ \partial}{r}}}\,{{H}_{2}}(\,{r}, {t}\,)\,
 \!  \right) \,(\,{r} - 2\,{m}\,) - (\, - 3\,{m} + 2\,{r}\,)\,
 \left( \! \,{\frac {{ \partial}}{{ \partial}{r}}}\,{\rm K}(\,{r}
, {t}\,)\, \!  \right)  -  \left( \! \,{\frac {{ \partial}}{{ 
\partial}{t}}}\,{{H}_{1}}(\,{r}, {t}\,)\, \!  \right) \,{r} \! 
\! \left. {\vrule height1.22em width0em depth1.22em} \right) 
{\rm u}(\,{ \theta}\,) \\
 & & \mbox{} +  \left( \! \, - \,{\displaystyle \frac {1}{2}}\,
{\displaystyle \frac {{\rm cos}(\,{ \theta}\,)\,{{H}_{0}}(\,{r}, 
{t}\,)}{{\rm sin}(\,{ \theta}\,)}} + {\displaystyle \frac {1}{2}}
\,{\displaystyle \frac {{\rm cos}(\,{ \theta}\,)\,{{H}_{2}}(\,{r}
, {t}\,)}{{\rm sin}(\,{ \theta}\,)}}\, \!  \right) \, \left( \! 
\,{\frac {{ \partial}}{{ \partial}{ \theta}}}\,{\rm u}(\,{ \theta
}\,)\, \!  \right) 
\end{eqnarray*}
\end{maplelatex}

\noindent Here we seem to run into trouble; our expression has both $u_i^\prime$ and $u_i$ terms. The projection $\langle u_i | 
u_j^\prime \rangle$ will in general be non-zero. Thus we could not decouple the modes if this expression were summed over. But when 
we combine this expression with $R_{\phi\phi}^{(1)}$ we will find a, seemingly fortuitous,  resolution. 

\bigskip

\begin{mapleinput}
g1:=kfactor(hcollect(coeff(collect(grcomponent(R(dn,dn),[phi,phi]),epsilon),epsilon,1)
,ufuncs,lperts),(cos(theta)-1)*(cos(theta)+1));
\end{mapleinput}
\begin{maplelatex}
\begin{eqnarray*}
\lefteqn{{\it g1} := (\,{\rm cos}(\,{ \theta}\,) + 1\,)\,(\,{\rm 
cos}(\,{ \theta}\,) - 1\,) \left( {\vrule 
height1.22em width0em depth1.22em} \right. \! \!  - \,
{\displaystyle \frac {1}{2}} \left( {\vrule 
height0.79em width0em depth0.79em} \right. \! \! 2\,{{H}_{2}}(\,{
r}, {t}\,)\,{r} - 4\,{{H}_{2}}(\,{r}, {t}\,)\,{m} - 2\,{\rm K}(\,
{r}, {t}\,)\,{r} + 4\,{\rm K}(\,{r}, {t}\,)\,{m}} \\
 & & \mbox{} + {\rm K}(\,{r}, {t}\,)\,{j}\,{r} - 2\,{\rm K}(\,{r}
, {t}\,)\,{j}\,{m} + {r}^{3}\, \left( \! \,{\frac {{ \partial}^{2
}}{{ \partial}{t}^{2}}}\,{\rm K}(\,{r}, {t}\,)\, \!  \right)  - {
r}^{3}\, \left( \! \,{\frac {{ \partial}^{2}}{{ \partial}{r}^{2}
}}\,{\rm K}(\,{r}, {t}\,)\, \!  \right)  \\
 & & \mbox{} + 4\,{r}^{2}\, \left( \! \,{\frac {{ \partial}^{2}}{
{ \partial}{r}^{2}}}\,{\rm K}(\,{r}, {t}\,)\, \!  \right) \,{m}
 - 4\,{r}\, \left( \! \,{\frac {{ \partial}^{2}}{{ \partial}{r}^{
2}}}\,{\rm K}(\,{r}, {t}\,)\, \!  \right) \,{m}^{2} +  \left( \! 
\,{\frac {{ \partial}}{{ \partial}{r}}}\,{{H}_{0}}(\,{r}, {t}\,)
\, \!  \right) \,{r}^{2} - 4\, \left( \! \,{\frac {{ \partial}}{{
 \partial}{r}}}\,{{H}_{0}}(\,{r}, {t}\,)\, \!  \right) \,{m}\,{r}
 \\
 & & \mbox{} + 4\, \left( \! \,{\frac {{ \partial}}{{ \partial}{r
}}}\,{{H}_{0}}(\,{r}, {t}\,)\, \!  \right) \,{m}^{2} +  \left( 
\! \,{\frac {{ \partial}}{{ \partial}{r}}}\,{{H}_{2}}(\,{r}, {t}
\,)\, \!  \right) \,{r}^{2} - 4\, \left( \! \,{\frac {{ \partial}
}{{ \partial}{r}}}\,{{H}_{2}}(\,{r}, {t}\,)\, \!  \right) \,{m}\,
{r} + 4\, \left( \! \,{\frac {{ \partial}}{{ \partial}{r}}}\,{{H}
_{2}}(\,{r}, {t}\,)\, \!  \right) \,{m}^{2} \\
 & & \mbox{} + 14\, \left( \! \,{\frac {{ \partial}}{{ \partial}{
r}}}\,{\rm K}(\,{r}, {t}\,)\, \!  \right) \,{m}\,{r} - 12\,
 \left( \! \,{\frac {{ \partial}}{{ \partial}{r}}}\,{\rm K}(\,{r}
, {t}\,)\, \!  \right) \,{m}^{2} - 4\, \left( \! \,{\frac {{ 
\partial}}{{ \partial}{r}}}\,{\rm K}(\,{r}, {t}\,)\, \!  \right) 
\,{r}^{2} - 2\, \left( \! \,{\frac {{ \partial}}{{ \partial}{t}}}
\,{{H}_{1}}(\,{r}, {t}\,)\, \!  \right) \,{r}^{2} \\
 & & \mbox{} + 4\, \left( \! \,{\frac {{ \partial}}{{ \partial}{t
}}}\,{{H}_{1}}(\,{r}, {t}\,)\, \!  \right) \,{r}\,{m} \! 
\! \left. {\vrule height0.79em width0em depth0.79em} \right) 
{\rm u}(\,{ \theta}\,) \left/ {\vrule 
height0.37em width0em depth0.37em} \right. \! \! (\,{r} - 2\,{m}
\,) \\
 & & \mbox{} - {\displaystyle \frac {1}{2}}\,{\displaystyle 
\frac {{\rm sin}(\,{ \theta}\,)\,{\rm cos}(\,{ \theta}\,)\,
 \left( \! \,{{H}_{2}}(\,{r}, {t}\,) - {{H}_{0}}(\,{r}, {t}\,)\,
 \!  \right) \, \left( \! \,{\frac {{ \partial}}{{ \partial}{ 
\theta}}}\,{\rm u}(\,{ \theta}\,)\, \!  \right) }{(\,{\rm cos}(\,
{ \theta}\,) - 1\,)\,(\,{\rm cos}(\,{ \theta}\,) + 1\,)}} \! 
\! \left. {\vrule height1.22em width0em depth1.22em} \right) 
\mbox{\hspace{156pt}}
\end{eqnarray*}
\end{maplelatex}

\noindent Here again we have the potential for mode coupling from the presence of both $u_i^\prime$ and $u_i$ terms. If, however, we 
add/subtract these last two components a remarkable thing happens:

\bigskip

\begin{mapleinput}
hcollect(a1+g1/sin(theta)^2,ufuncs,lperts);
\end{mapleinput}
\begin{maplelatex}
\begin{eqnarray*}
\lefteqn{ \left( {\vrule height1.22em width0em depth1.22em}
 \right. \! \! {\displaystyle \frac {1}{2}}\,(\,{j} + 4\,)\,{{H}
_{2}}(\,{r}, {t}\,) + (\, - 2 + {j}\,)\,{\rm K}(\,{r}, {t}\,) + 
{\displaystyle \frac {{r}^{3}\, \left( \! \,{\frac {{ \partial}^{
2}}{{ \partial}{t}^{2}}}\,{\rm K}(\,{r}, {t}\,)\, \!  \right) }{{
r} - 2\,{m}}} -  \left( \! \,{\frac {{ \partial}^{2}}{{ \partial}
{r}^{2}}}\,{\rm K}(\,{r}, {t}\,)\, \!  \right) \,(\,{r} - 2\,{m}
\,)\,{r}} \\
 & & \mbox{} - {\displaystyle \frac {1}{2}}\,{{H}_{0}}(\,{r}, {t}
\,)\,{j} +  \left( \! \,{\frac {{ \partial}}{{ \partial}{r}}}\,{{
H}_{0}}(\,{r}, {t}\,)\, \!  \right) \,(\,{r} - 2\,{m}\,) + 
 \left( \! \,{\frac {{ \partial}}{{ \partial}{r}}}\,{{H}_{2}}(\,{
r}, {t}\,)\, \!  \right) \,(\,{r} - 2\,{m}\,) \\
 & & \mbox{} - 2\,(\, - 3\,{m} + 2\,{r}\,)\, \left( \! \,{\frac {
{ \partial}}{{ \partial}{r}}}\,{\rm K}(\,{r}, {t}\,)\, \! 
 \right)  - 2\, \left( \! \,{\frac {{ \partial}}{{ \partial}{t}}}
\,{{H}_{1}}(\,{r}, {t}\,)\, \!  \right) \,{r} \! \! \left. 
{\vrule height1.22em width0em depth1.22em} \right) {\rm u}(\,{ 
\theta}\,)\mbox{\hspace{100pt}}
\end{eqnarray*}
\end{maplelatex}

\noindent By adding these two expressions we have eliminated the $u_i^\prime$ terms, and can therefore select any ``$i^{th}$'' term by 
simply projecting with $u_i$. Similarly:

\bigskip

\begin{mapleinput}
factor(hcollect(a1-g1/sin(theta)^2,ufuncs,lperts));
\end{mapleinput}
\begin{maplelatex}
\[
{\displaystyle \frac {1}{2}}\,{\displaystyle \frac { \left( \! \,
{{H}_{2}}(\,{r}, {t}\,) - {{H}_{0}}(\,{r}, {t}\,)\, \!  \right) 
\, \left( \! \,{j}\,{\rm u}(\,{ \theta}\,)\,{\rm sin}(\,{ \theta}
\,) + 2\,{\rm cos}(\,{ \theta}\,)\, \left( \! \,{\frac {{ 
\partial}}{{ \partial}{ \theta}}}\,{\rm u}(\,{ \theta}\,)\, \! 
 \right) \, \!  \right) }{{\rm sin}(\,{ \theta}\,)}}
\]
\end{maplelatex}

\noindent While this subtraction seems to again leave us with the potential for mode coupling, we notice that:
$$
i(i+1)u_i + 2\cot(\theta)u_i^\prime(\theta) = -(1-x^2)d^2 u_i/dx^2 ~~, 
$$
where $x=\cos(\theta)$. Recognizing this as the $P^2_i (x)$ associated Legendre function, and recalling that these also form 
a complete orthogonal set, 
we can, therefore, uniquely select out any term in such a series by projecting with: 
$$
 i(i+1)u_i + 2\cot(\theta)d u_i/d\theta~.
$$

The decoupling of the linear problem is thus reduced to a prescription for performing seven projections,
$$
\left\langle \left. F_i\left(R_{\alpha\beta}^{(1)}\right) \right| G_i\left(P_l\right)\right\rangle = 0~,
$$
from which one obtains seven PDEs that govern the MP functions. Here $i=1..7$, $F_i$ produces a linear combination of its 
arguments, and $G_i$ is at most a function of the given argument and its first derivative. We can summarize the details of this decoupling 
procedure in the following table that relates the seven linear combinations of the perturbed Ricci components to their associated 
projection functions.
\begin{center}
\begin{tabular}{|c|c|} \hline
Perturbed Ricci components & $l$-mode Projection function \\ \hline \hline
$\delta R_{r\theta}$~,~$\delta R_{\theta t}$ & $ dP_l/d\theta$ \\ \hline
$\delta R_{rr}$~,~$\delta R_{rt}$~,~$\delta R_{tt}$ & $P_l$ \\ \hline
$\delta R_{\theta\theta}+\delta R_{\phi\phi}/\sin^2(\theta)$ & $P_l$ \\ \hline
$\delta R_{\theta\theta}-\delta R_{\phi\phi}/\sin^2(\theta)$ & $l(l+1)P_l+2\cot(\theta)dP_l/d\theta$ \\ \hline
\end{tabular}

\medskip
Table 1.

\end{center}
This table is the central result of this section, and is a kind of Rosetta
stone for the Schwarzschild perturbation analysis. It
allows one to consistently and algorithmically decouple the angular perturbations,
regardless of the order of the perturbation expansion\footnote{
This is a direct result of the fact that higher order perturbations will simply
produce the same seven equations, now in terms of the higher order functions, 
along with
possible source terms.}. Thus the $\delta R_{\alpha\beta}$ can be any order of
the perturbed Ricci tensor, $R_{\alpha\beta}^{(n)}$, or even the total
perturbation, $\sum_n\epsilon^n R_{\alpha\beta}^{(n)}$.

While we have managed to decouple the perturbation modes,
we should note that there is a great redundancy in the system of seven PDEs
that are generated by these projections. The remainder of the solution to
the linear problem is, essentially, the elimination of this excess information.
As a simple example of this redundancy, and as a prelude to the results of the
next section, we note that the
``$\theta\theta-\phi\phi$'' projection in the linear analysis gives,
$$
\int_0^\pi d\theta \sin(\theta) P_l^2(\theta)\left( 
\delta R_{\theta\theta}-\delta R_{\phi\phi}/\sin^2(\theta)\right)=0
\longrightarrow H_0(r,t)=H_2(r,t)~.
$$
At the linear level then, $H_0$ and $H_2$ are degenerate\footnote{We will find
that this degeneracy is broken at the second order level by the presence of
``source terms''.}. 

The reduction of the system is typically accomplished by
making the Zerilli transformation,
$$
H_1(r,t)=\frac{(2r^2-6rm-3m^2)}{(r-2m)(2r+3m)}\chi(r,t)+
\frac{r^2\eta(r,t)}{r-2m}
~~,~~
\frac{\partial K(r,t)}{\partial t}=
6\frac{(r^2+rm+m^2)}{r^2(2r+3m)}\chi(r,t)+\eta(r,t)~.
$$
With these substitutions the redundancy can now be expressed as:
$\eta(r,t)=(1-2m/r)\partial_r\chi$. Using this result to eliminate $\eta$
throughout the system one then finds 
a ``wave equation'' in the single new function $\chi(r,t)$. This is the 
celebrated Zerilli ``wave equation'':
$$
\frac{\partial^2\chi(r,t)}{\partial t^2}-\frac{\partial^2 \chi(r,t)}{\partial {r^*}^2} + V_l(r)\chi(r,t)=0~.
$$
Here, $r^*=r+2M \log (r/2M-1)$, and the effective potential, $V$, has an explicit dependence on the angular mode number, $l$. 
Considering our discussion of perturbation theory from the last section, we expect that the second order Zerilli wave equation will be the 
same as that of the linear case, but with the addition of a new source term. Finding the form of this source term, which is not 
unique\footnote{See, for example, \cite{GL}, for a discussion of the gauge dependence of the source term.}, is the central problem of 
the second order calculation. Calculating the source term will, ultimately, be the goal of the next section.   

As the final comment of this section, we invite the interested reader to 
download the Maple worksheet that produced the linearized results presented
here. One is then free to reproduce these results and experiment with the
tools at hand. In particular one should try to replicate the linearized 
equations without the aid of the {\bf termsimp()} and {\bf hcollect()}
routines.

\subsection*{The Quadratic Schwarzschild Analysis}

We found in the last section that the non-trivial components of the linearly perturbed Ricci tensor grouped into five 
components, and two linear combinations of components, each having an associated projection function. Making use of this linear 
analysis we can now solve the second order Schwarzschild problem. Here the second order perturbations appear exactly as the linear 
ones have, but with the addition of ``source'' terms that are quadratic in the (assumed known) linear perturbations. The key here is that 
while the quadratic source terms will in general be a combination of an arbitrary number of multipoles, we now have a unique 
prescription for determining their contribution to any desired second order multipole perturbation. 

From the point of view of the second order analysis, the form of the linear perturbation functions are, apart from consistency, simply a 
matter of choice. The recent work that has appeared in the literature assumes that the linear perturbations have only a quadrapole 
component, see, $e.g.$ \cite{GL}. This restriction can be justified on physical ground by arguing that most of the spectral power will be 
concentrated in this mode and that the full source could be reconstructed by simply summing over the actual multipole modes of the 
linear problem. But perhaps the best rational for this restriction is that there exists a particular class of interesting problems, see, $e.g.$ \cite{GL},
for which the only non-zero multipole is, at the linear level, the quadrapole.
For this case then, the treatment is exact, and it is
this scenario that we will study.
The remainder of this section will outline the calculation of the second 
order quadrapole results, $i.e.$, ${g^{(1)}_{\alpha\beta}}$, and,
${g^{(2)}_{\alpha\beta}}$, will both be pure quadrapole moments. In solving
this particular problem then, we are examining how the first order quadrapole
moment seeds the second order quadrapole moment. We will return to discuss this
point in our conclusion, but for now one should simply notice that the 
following analysis could easily be repeated with {\em any} two given multipoles.
Further details of the calculation can be found in Appendix B, and as in the 
last section, the complete Maple worksheet that produced these results can be
downloaded from the GRTensor website.

The quadratically perturbed covariant metric for the ``quadrapole-quadrapole''
perturbation is a simple extension of the linear case.

\bigskip

\begin{maplelatex}
\begin{eqnarray*}
\lefteqn{{\it \:ds}^{2}={\displaystyle \frac { \left( \! \,1 + 
{\displaystyle \frac {1}{2}}\, \left( \! \,{ \varepsilon}\,{{h}_{
2}}(\,{r}, {t}\,) + { \varepsilon}^{2}\,{{H}_{2}}(\,{r}, {t}\,)\,
 \!  \right) \, \left( \! \,3\,{\rm cos}(\,{ \theta}\,)^{2} - 1\,
 \!  \right) \, \!  \right) \,{\it \:d}\,{r}^{{\it 2\:}}}{1 - 2\,
{\displaystyle \frac {{m}}{{r}}}}}} \\
 & & \mbox{} +  \left( \! \,{ \varepsilon}\,{{h}_{1}}(\,{r}, {t}
\,) + { \varepsilon}^{2}\,{{H}_{1}}(\,{r}, {t}\,)\, \!  \right) 
\, \left( \! \,3\,{\rm cos}(\,{ \theta}\,)^{2} - 1\, \!  \right) 
\,{\it \:d}\,{r}^{{\:}}\,{\it d\:}\,{t}^{{\:}} \\
 & & \mbox{} + {r}^{2}\, \left( \! \,1 + {\displaystyle \frac {1
}{2}}\, \left( \! \,{ \varepsilon}\,{\rm k}(\,{r}, {t}\,) + { 
\varepsilon}^{2}\,{\rm K}(\,{r}, {t}\,)\, \!  \right) \, \left( 
\! \,3\,{\rm cos}(\,{ \theta}\,)^{2} - 1\, \!  \right) \, \! 
 \right) \,{\it \:d}\,{ \theta}^{{\it 2\:}} \\
 & & \mbox{} + {r}^{2}\,{\rm sin}(\,{ \theta}\,)^{2}\, \left( \! 
\,1 + {\displaystyle \frac {1}{2}}\, \left( \! \,{ \varepsilon}\,
{\rm k}(\,{r}, {t}\,) + { \varepsilon}^{2}\,{\rm K}(\,{r}, {t}\,)
\, \!  \right) \, \left( \! \,3\,{\rm cos}(\,{ \theta}\,)^{2} - 1
\, \!  \right) \, \!  \right) \,{\it \:d}\,{ \phi}^{{\it 2\:}} \\
 & & \mbox{} -  \left( \! \,1 - 2\,{\displaystyle \frac {{m}}{{r}
}}\, \!  \right) \, \left( \! \,1 - {\displaystyle \frac {1}{2}}
\, \left( \! \,{ \varepsilon}\,{{h}_{0}}(\,{r}, {t}\,) + { 
\varepsilon}^{2}\,{{H}_{0}}(\,{r}, {t}\,)\, \!  \right) \,
 \left( \! \,3\,{\rm cos}(\,{ \theta}\,)^{2} - 1\, \!  \right) \,
 \!  \right) \,{\it \:d}\,{t}^{{\it 2\:}}
\end{eqnarray*}
\end{maplelatex}
\begin{maplelatex}
\[
{\it Constraints}= \left[ \! \,{ \varepsilon}^{4}=0, { 
\varepsilon}^{5}=0, { \varepsilon}^{6}=0, { \varepsilon}^{7}=0, {
 \varepsilon}^{8}=0, { \varepsilon}^{9}=0, { \varepsilon}^{10}=0
, { \varepsilon}^{3}=0\, \!  \right] 
\]
\end{maplelatex}

\bigskip

\noindent As can be seen from the form of the metric, the lower case functions are the first order perturbations, and the upper case 
functions are the second order perturbations.

The calculation now proceeds exactly as in the linear case, except that we now
expand the contravariant metric to second order and truncate higher than 
$O(\epsilon^2)$ terms during the calculation of the Ricci tensor. 
One can gain an immediate appreciation for the
increase in complexity of this problem over the linear case by examining the
size of the second order Ricci tensor.

\bigskip

\begin{maplelatex}
\[
{{\it R\:}_{{r}}}\,{{}_{{r}}}={\it \:86717\:words.\:Exceeds\:
grOptionDisplayLimit}
\]
\end{maplelatex}
\begin{maplelatex}
\[
{{\it R\:}_{{r}}}\,{{}_{{ \theta}}}={\it \:28579\:words.\:Exceeds
\:grOptionDisplayLimit}
\]
\end{maplelatex}
\begin{maplelatex}
\[
{{\it R\:}_{{r}}}\,{{}_{{t}}}={\it \:53004\:words.\:Exceeds\:
grOptionDisplayLimit}
\]
\end{maplelatex}
\begin{maplelatex}
\[
{{\it R\:}_{{ \theta}}}\,{{}_{{ \theta}}}={\it \:87057\:words.\:
Exceeds\:grOptionDisplayLimit}
\]
\end{maplelatex}
\begin{maplelatex}
\[
{{\it R\:}_{{ \theta}}}\,{{}_{{t}}}={\it \:20438\:words.\:Exceeds
\:grOptionDisplayLimit}
\]
\end{maplelatex}
\begin{maplelatex}
\[
{{\it R\:}_{{ \phi}}}\,{{}_{{ \phi}}}={\it \:72053\:words.\:
Exceeds\:grOptionDisplayLimit}
\]
\end{maplelatex}
\begin{maplelatex}
\[
{{\it R\:}_{{t}}}\,{{}_{{t}}}={\it \:82672\:words.\:Exceeds\:
grOptionDisplayLimit}
\]
\end{maplelatex}

\noindent As one might expect, the quadratically perturbed Ricci tensor is
substantially larger than its linear counterpart, and it is here that our new
simplification routines will become indispensable.
Of course, 
what we have really calculated here is,
$$
R_{\alpha\beta}\simeq  R^{(0)}_{\alpha\beta} + \epsilon R^{(1)}_{\alpha\beta} + \epsilon^2 R^{(2)}_{\alpha\beta}~,
$$
where $R^{(0)}_{\alpha\beta}=0$. Given that the Ricci tensor now contains terms that can be either linear or quadratic in $\epsilon$, 
care must be taken when extracting a particular order of perturbation.

In order to determine the PDEs that govern the second order perturbations
we simply follow the procedure set out in the linear analysis. Here we select
only the $\epsilon^2$ components of the Ricci tensor, and we perform the $l=2$
projections. This projects the squared quadrapole of the linear perturbation
onto the second order quadrapole\footnote{In the case of a more general 
problem, $i.e.$, one that had more than a single second order perturbation, this
procedure
would also decouple the second order multipoles and accomplish the associated
projections.}.
The second order perturbation functions will, necessarily, appear in these
PDEs in exactly 
the same manner as the linear perturbation functions appear in the last 
section. The only new feature, which lies at the heart of 
the second order theory, is the manner in which the quadratically occuring linear perturbations appear. We will not show the raw 
expressions for most of these PDEs as many of their source terms are
quite large. Instead, we give the detailed expressions 
for two of the simpler equations, both to illustrate how these equations can
differ
from their linear counterparts and to make contact with the results presented
in \cite{GL}. The results of the 
``$\theta\theta-\phi\phi$'' and 
``$t\theta$'' projections yield,
\begin{maplelatex}
\[ \! \,
{{H}_{2}}(\,{r}, {t}\,)={{H}_{0}}(\,{r}, {t}\,) - 
{\displaystyle \frac {1}{7}}\,{{h}_{0}}(\,{r}, {t}\,)^{2} + 
{\displaystyle \frac {1}{7}}\,{{h}_{1}}(\,{r}, {t}\,)^{2}~,
\, \! \]
\end{maplelatex}
and,
\begin{maplelatex}
\begin{eqnarray*}
\lefteqn{2\,{\displaystyle \frac {{m}\,{{H}_{1}}(\,{r}, {t}\,)}{{
r}^{2}}} + {\displaystyle \frac {(\,{r} - 2\,{m}\,)\, \left( \! 
\,{\frac {{ \partial}}{{ \partial}{r}}}\,{{H}_{1}}(\,{r}, {t}\,)
\, \!  \right) }{{r}}} -  \left( \! \,{\frac {{ \partial}}{{ 
\partial}{t}}}\,{{H}_{0}}(\,{r}, {t}\,)\, \!  \right)  -  \left( 
\! \,{\frac {{ \partial}}{{ \partial}{t}}}\,{\rm K}(\,{r}, {t}\,)
\, \!  \right)  + {\displaystyle \frac {3}{14}}\, \left( \! \,
{\frac {{ \partial}}{{ \partial}{t}}}\,{{h}_{0}}(\,{r}, {t}\,)^{2
}\, \!  \right) } \\
 & & \mbox{} + {\displaystyle \frac {1}{7}}\, \left( \! \,
{\frac {{ \partial}}{{ \partial}{t}}}\,{\rm k}(\,{r}, {t}\,)^{2}
\, \!  \right)  - {\displaystyle \frac {3}{14}}\, \left( \! \,
{\frac {{ \partial}}{{ \partial}{t}}}\,{{h}_{1}}(\,{r}, {t}\,)^{2
}\, \!  \right) = 0~, \mbox{\hspace{218pt}}
\end{eqnarray*}
\end{maplelatex}

\noindent respectively. The degeneracy between $H_0$ and $H_2$ is therefore 
broken at second order by the appearance of ``source'' terms. Of course, one
can still eliminate $H_2$ from the analysis using this first of these 
expressions. As with the linearized case, the second order perturbations form
a very redundant system. The effective source terms, of course, inherit all of
the redundancy of the linear problem as well. The completion of our second 
order calculation will be the elimination of this redundancy, $\acute{a}$ $la$
Zerilli, in a manner that will allow us to reproduce the results of \cite{GL}.

The derivation of the first of the Zerilli equations proceeds by eliminating 
$\dot H_0$ from $R_{rr}^{(2)}$ using the $R_{rt}^{(2)}$ component of the Ricci
tensor. One must then eliminate the redundant source terms, a procedure that
is not unique. Here we choose to preferentially eliminate certain source
terms to reproduce the results in \cite{GL}. The details of this process are
not very enlightening and are relegated to the Maple worksheet that can be found
at the GRTensor website. The result of these manipulations is essentially the
raw result for the $\eta(r,t)$ expression which can be found in Appendix B.
One should note that the quadratic nature of the linear
perturbations that appear in these types of expressions make them almost
impossible to simplify with the standard Maple routines. To overcome this 
problem we use our new {\bf mcollect()} routine (see Appendix A),
after successive application of which, we find,
\begin{maplelatex}
\begin{eqnarray*}
\lefteqn{ {\vrule 
height1.22em width0em depth1.22em} \! \! { \eta}(\,{r}, {
t}\,)={\displaystyle \frac {(\,{r} - 2\,{m}\,)\, \left( \! \,
{\frac {{ \partial}}{{ \partial}{r}}}\,{ \chi}(\,{r}, {t}\,)\,
 \!  \right) }{{r}}} - {\displaystyle \frac {1}{7}}(\,{r} - 2\,{m
}\,) \left( {\vrule height1.22em width0em depth1.22em}
 \right. \! \!  - {{h}_{0}}(\,{r}, {t}\,)\, \left( \! \,{\frac {{
 \partial}}{{ \partial}{t}}}\,{{h}_{0}}(\,{r}, {t}\,)\, \! 
 \right) } \\
 & & \mbox{} + 2\, \left( \! \,{\frac {{ \partial}}{{ \partial}{t
}}}\,{{h}_{0}}(\,{r}, {t}\,)\, \!  \right) \,{\rm k}(\,{r}, {t}\,
) + {r}\, \left( \! \,{\frac {{ \partial}}{{ \partial}{r}}}\,
{\rm k}(\,{r}, {t}\,)\, \!  \right) \, \left( \! \,{\frac {{ 
\partial}}{{ \partial}{t}}}\,{{h}_{0}}(\,{r}, {t}\,)\, \! 
 \right)  + 2\,{\rm k}(\,{r}, {t}\,)\, \left( \! \,{\frac {{ 
\partial}}{{ \partial}{t}}}\,{\rm k}(\,{r}, {t}\,)\, \!  \right) 
 \\
 & & \mbox{} + 4\,{\displaystyle \frac {{{h}_{1}}(\,{r}, {t}\,)\,
{\rm k}(\,{r}, {t}\,)}{{r}}} - 2\,{\displaystyle \frac {{m}\,{{h}
_{1}}(\,{r}, {t}\,)\, \left( \! \,{\frac {{ \partial}}{{ \partial
}{r}}}\,{\rm k}(\,{r}, {t}\,)\, \!  \right) }{{r}}} + {{h}_{1}}(
\,{r}, {t}\,)\, \left( \! \,{\frac {{ \partial}}{{ \partial}{t}}}
\,{{h}_{1}}(\,{r}, {t}\,)\, \!  \right)  \\
 & & \mbox{} + 2\,{\displaystyle \frac {(\,{r} - 2\,{m}\,)\,{{h}
_{1}}(\,{r}, {t}\,)\,{{h}_{0}}(\,{r}, {t}\,)}{{r}^{2}}} - 
{\displaystyle \frac {{r}^{2}\, \left( \! \,{\frac {{ \partial}}{
{ \partial}{t}}}\,{\rm k}(\,{r}, {t}\,)\, \!  \right) \, \left( 
\! \,{\frac {{ \partial}}{{ \partial}{t}}}\,{{h}_{1}}(\,{r}, {t}
\,)\, \!  \right) }{{r} - 2\,{m}}} \\
 & & \mbox{} + 2\,{\displaystyle \frac { \left( \! \,{\frac {{ 
\partial}}{{ \partial}{t}}}\,{\rm k}(\,{r}, {t}\,)\, \!  \right) 
\,{{h}_{0}}(\,{r}, {t}\,)\,{m}}{{r} - 2\,{m}}} \! \! \left. 
{\vrule height1.22em width0em depth1.22em} \right)  \left/ 
{\vrule height0.37em width0em depth0.37em} \right. \! \! (\,2\,{r
} + 3\,{m}\,)~. \! \! {\vrule 
height1.22em width0em depth1.22em} 
\end{eqnarray*}
\end{maplelatex}

\noindent This result, which appears here exactly as outputed from the Maple
session,
agrees with the result found in [1]. The new source terms that appear in this
second order result add a considerable degree of complexity to the problem, as
one must use this result to eliminate all of the $\eta$ dependence, which 
includes up to third order derivatives, from the analysis.
One should also note that without the use of
our new simplification routines this analysis would not have 
been as seamless: Ideally, when performing computer algebra calculations, one
should never have to resort to a ``by hand'' calculation 
as this increases the chance of error by a $very$ large factor. Our new
simplification routines were designed precisely to avoid the 
necessity of {\em any} ``by hand'' calculations. 

The development of the second order Zerilli ``wave equation'' follows a similar
procedure to this last result. We first 
take the time derivative of $R_{r\theta}^{(2)}$, eliminate $\dot H_0$ and
$\dot H_0^\prime$ using $R_{rt}^{(2)}$, make the Zerilli substitutions, and 
substitute for $\eta$ from our first Zerilli result. We then repeat this 
process for $R_{t\theta}^{(2)}$ and add the resulting equations such that the 
$\ddot \chi^\prime$ term vanishes. The result of these operations
is essentially the raw form 
of the result we are seeking. Having $421$ terms, this expression is, however, 
rather large.

In order to make contact with the results presented in \cite{GL}, we now
work with the ``renormalized'' perturbation function,
$\zeta$, defined by,

\medskip

\begin{maplelatex}
\[
 \! \,{ \chi}(\,{r}, {t}\,)={ \zeta}(\,{
r}, {t}\,) + {\displaystyle \frac {2}{7}}\,{\displaystyle \frac {
{r}^{2}\,{\rm k}(\,{r}, {t}\,)\, \left( \! \,{\frac {{ \partial}
}{{ \partial}{t}}}\,{\rm k}(\,{r}, {t}\,)\, \!  \right) }{2\,{r}
 + 3\,{m}}} + {\displaystyle \frac {2}{7}}\,{\rm k}(\,{r}, {t}\,)
 ^{2}\, \!  
 \]
 \end{maplelatex}

\noindent This choice of transformation reduces the number of terms 
by over a factor of three. The next problem we must 
tackle is to eliminate all of the redundant information contained in the source 
terms for $\zeta$. In principle this should be relatively 
simple, and certainly straight forward; in practice it is neither. The basic 
necessity is to use all of the PDEs found in the linear 
analysis to eliminate as many of the ``higher order'' derivatives as is
possible, and to preferentially eliminate any $h_1$ dependence. The 
rational for this is that, to linear order, all of the radiation information is 
determined by the Zerilli function, $\psi$, which depends only 
on $k$ and $h_0$. One therefore expects that the source terms can be expressed
solely in terms of these two functions, and ultimately, 
solely in terms of $\psi$. Our approach is thus to express $h_0$ in terms of 
$\psi$ and $k$ once the $h_1$ dependence has been 
eliminated from the source terms. 

\begin{maplelatex}
\[
  \! \,{{h}_{0}}(\,{r}, {t}\,)=3\,{\displaystyle \frac {
  \left( \! \,{ \psi}(\,{r}, {t}\,) - {\displaystyle \frac {1}{3}}
  \,{r}\,{\rm k}(\,{r}, {t}\,)\, \!  \right) \,(\,2\,{r} + 3\,{m}\,
  )}{{r}\,(\,{r} - 2\,{m}\,)}} +  \left( \! \,{\frac {{ \partial}}{
  { \partial}{r}}}\,{\rm k}(\,{r}, {t}\,)\, \!  \right) \,{r}\, \! 
   \]
   \end{maplelatex}

\noindent After performing all of these eliminations\footnote{These 
details can be found in the worksheet at the GRTensor website.}, 
making the above 
substitution for $h_0$, and, further simplifying the result to eliminate the
$k$ dependence, we arrive at the unsimplified form of our 
final result, which is still rather large at $154$ terms. To simplify this
expression we once again use a repeated application of {\bf mcollect()}. 
Here we must
collect and simplify the 17 different types of quadratic $\psi$ terms that
occur. 
Making one last set of substitutions, $\mu=(r-2m)$, and, $\lambda=(2r+3m)$,
and extracting the source terms from the resulting expression yields:

\bigskip

\begin{maplelatex}
\begin{eqnarray*}
\lefteqn{ {\displaystyle \frac {12}{7}}{ \mu}
^{3} \left( {\vrule height1.36em width0em depth1.36em}
 \right. \! \!  - 12\,{\displaystyle \frac {(\,{r}^{2} + {m}\,{r}
 + {m}^{2}\,)^{2}\, \left( \! \,{\frac {{ \partial}}{{ \partial}{
t}}}\,{ \psi}(\,{r}, {t}\,)\, \!  \right) ^{2}}{{r}^{4}\,{ \mu}^{
3}\,{ \lambda}}}} \\
 & & \mbox{} - 4\,{\displaystyle \frac {(\,2\,{r}^{3} + 4\,{r}^{2
}\,{m} + 9\,{r}\,{m}^{2} + 6\,{m}^{3}\,)\,{ \psi}(\,{r}, {t}\,)\,
{\rm \%2}}{{r}^{6}\,{ \lambda}}} +  \\
 & & (\,112\,{r}^{5} + 480\,{r}^{4}\,{m} + 692\,{r}^{3}\,{m}^{2}
 + 762\,{r}^{2}\,{m}^{3} + 441\,{r}\,{m}^{4} + 144\,{m}^{5}\,)\,{
 \psi}(\,{r}, {t}\,)\, \left( \! \,{\frac {{ \partial}}{{ 
\partial}{t}}}\,{ \psi}(\,{r}, {t}\,)\, \!  \right)  \\
 & &  \left/ {\vrule height0.43em width0em depth0.43em}
 \right. \! \!  \left( \! \,{r}^{5}\,{ \mu}^{2}\,{ \lambda}^{3}\,
 \!  \right) \mbox{} - {\displaystyle \frac {1}{3}}\,
{\displaystyle \frac { \left( \! \,{\frac {{ \partial}}{{ 
\partial}{t}}}\,{ \psi}(\,{r}, {t}\,)\, \!  \right) \, \left( \! 
\,{\frac {{ \partial}^{3}}{{ \partial}{r}^{3}}}\,{ \psi}(\,{r}, {
t}\,)\, \!  \right) }{{r}^{2}}} \\
 & & \mbox{} + {\displaystyle \frac {1}{3}}\,{\displaystyle 
\frac {(\,18\,{r}^{3} - 4\,{r}^{2}\,{m} - 33\,{r}\,{m}^{2} - 48\,
{m}^{3}\,)\, \left( \! \,{\frac {{ \partial}}{{ \partial}{r}}}\,{
 \psi}(\,{r}, {t}\,)\, \!  \right) \, \left( \! \,{\frac {{ 
\partial}}{{ \partial}{t}}}\,{ \psi}(\,{r}, {t}\,)\, \!  \right) 
}{{r}^{4}\,{ \mu}^{2}\,{ \lambda}}} \\
 & & \mbox{} + {\displaystyle \frac {1}{3}}\,{\displaystyle 
\frac {(\,12\,{r}^{3} + 36\,{r}^{2}\,{m} + 59\,{r}\,{m}^{2} + 90
\,{m}^{3}\,)\, \left( \! \,{\frac {{ \partial}}{{ \partial}{r}}}
\,{ \psi}(\,{r}, {t}\,)\, \!  \right) ^{2}}{{r}^{6}\,{ \mu}}} \\
 & & \mbox{} + 12\,{\displaystyle \frac {(\,2\,{r}^{5} + 9\,{r}^{
4}\,{m} + 6\,{r}^{3}\,{m}^{2} - 2\,{r}^{2}\,{m}^{3} - 15\,{r}\,{m
}^{4} - 15\,{m}^{5}\,)\,{ \psi}(\,{r}, {t}\,)^{2}}{{r}^{8}\,{ \mu
}^{2}\,{ \lambda}}} \\
 & & \mbox{} - 4\,{\displaystyle \frac {(\,{r}^{2} + {m}\,{r} + {
m}^{2}\,)\, \left( \! \,{\frac {{ \partial}}{{ \partial}{t}}}\,{ 
\psi}(\,{r}, {t}\,)\, \!  \right) \,{\rm \%1}}{{r}^{3}\,{ \mu}^{2
}}} - 2 \\
 & & (\,32\,{r}^{5} + 88\,{r}^{4}\,{m} + 296\,{r}^{3}\,{m}^{2} + 
510\,{r}^{2}\,{m}^{3} + 561\,{r}\,{m}^{4} + 270\,{m}^{5}\,)\,{ 
\psi}(\,{r}, {t}\,)\, \left( \! \,{\frac {{ \partial}}{{ \partial
}{r}}}\,{ \psi}(\,{r}, {t}\,)\, \!  \right)  \left/ {\vrule 
height0.43em width0em depth0.43em} \right. \! \!  \\
 & &  \left( \! \,{r}^{7}\,{ \mu}\,{ \lambda}^{2}\, \!  \right) 
\mbox{} + {\displaystyle \frac {1}{3}}\,{\displaystyle \frac {
 \left( \! \,{\frac {{ \partial}}{{ \partial}{r}}}\,{ \psi}(\,{r}
, {t}\,)\, \!  \right) \, \left( \! \,{\frac {{ \partial}^{3}}{{ 
\partial}{t}\,{ \partial}{r}^{2}}}\,{ \psi}(\,{r}, {t}\,)\, \! 
 \right) }{{r}^{2}}} - {\displaystyle \frac {(\,2\,{r}^{2} - {m}
^{2}\,)\, \left( \! \,{\frac {{ \partial}}{{ \partial}{t}}}\,{ 
\psi}(\,{r}, {t}\,)\, \!  \right) \,{\rm \%2}}{{r}^{3}\,{ \mu}\,{
 \lambda}}} \\
 & & \mbox{} + {\displaystyle \frac {(\,8\,{r}^{2} + 12\,{m}\,{r}
 + 7\,{m}^{2}\,)\,{ \psi}(\,{r}, {t}\,)\,{\rm \%1}}{{r}^{4}\,{ 
\mu}\,{ \lambda}}} + {\displaystyle \frac {1}{3}}\,
{\displaystyle \frac {(\,3\,{r} - 7\,{m}\,)\, \left( \! \,
{\frac {{ \partial}}{{ \partial}{r}}}\,{ \psi}(\,{r}, {t}\,)\,
 \!  \right) \,{\rm \%1}}{{r}^{3}\,{ \mu}}} \\
 & & \mbox{} - {\displaystyle \frac {{m}\,{ \psi}(\,{r}, {t}\,)\,
 \left( \! \,{\frac {{ \partial}^{3}}{{ \partial}{t}\,{ \partial}
{r}^{2}}}\,{ \psi}(\,{r}, {t}\,)\, \!  \right) }{{r}^{3}\,{ 
\lambda}}} + {\displaystyle \frac {4}{3}}\,{\displaystyle \frac {
(\,3\,{r}^{2} + 5\,{m}\,{r} + 6\,{m}^{2}\,)\, \left( \! \,
{\frac {{ \partial}}{{ \partial}{r}}}\,{ \psi}(\,{r}, {t}\,)\,
 \!  \right) \,{\rm \%2}}{{r}^{5}}} + {\displaystyle \frac {1}{3
}}\,{\displaystyle \frac {{ \mu}\,{ \lambda}\,{\rm \%2}^{2}}{{r}
^{4}}} \\
 & & \mbox{} - {\displaystyle \frac {1}{3}}\,{\displaystyle 
\frac {{ \lambda}\,{\rm \%1}^{2}}{{r}^{2}\,{ \mu}}} \! \! \left. 
{\vrule height1.36em width0em depth1.36em} \right)  \left/ 
{\vrule height0.37em width0em depth0.37em} \right. \! \! { 
\lambda} \\
 & & {\rm \%1} := {\frac {{ \partial}^{2}}{{ \partial}{t}\,{ 
\partial}{r}}}\,{ \psi}(\,{r}, {t}\,) \\
 & & {\rm \%2} := {\frac {{ \partial}^{2}}{{ \partial}{r}^{2}}}\,
{ \psi}(\,{r}, {t}\,)
\end{eqnarray*}
\end{maplelatex}

\noindent This is our final result, and also appears exactly as outputed from 
the Maple worksheet. We have derived the effective Zerilli source term
that dictates how the linear quadrapole perturbation seeds the second order 
quadrapole perturbation for a Schwarzschild black hole. This expression is 
almost identical to that presented in \cite{GL}.
The difference between the above expression and the previously published
result is
that the $\dot \psi^2$ 
term that appears in \cite{GL} only has factor of $1/\mu^2$, while we have
found that it has a $1/\mu^3$ dependence.

To conclude this section we again invite the interested reader to download the
full Maple worksheet from the GRTensor website. One can then reproduce these
calculations and, of course, experiment. One could, for instance, alter 
the worksheet to calculate the source term for any particular pair of 
linear-quadratic multipole perturbations.



\bigskip

\subsection*{Conclusion and Discussion}

This article has presented a detailed account of our calculation of second order perturbations to a Schwarzschild black hole. Our 
methodology was chosen to illustrate both the utility, and necessity, of employing computer algebraic methods to examine these types of 
problems. We began our analysis with an explicit demonstration that the linear Schwarzschild perturbations decouple, a result that is 
well known but that is not always clearly presented. We then used these linear
results to examine the second order pure quadrapole case. 
These second 
order results confirm, with one minor exception\footnote{One, out of seventeen,
of our Zerilli source terms differs from the previously published results.}, 
the earlier work by Gleiser, {\em et. al.}, 
in \cite{GL}.

As mentioned earlier, there are interesting problems for which the quadrapole
perturbation provides an exact description of the system 
to first order. Given this, one may then wish to consider how the linear
quadrapole seeds an arbitrary second order multipole 
perturbation. This more general result has been presented by Pullin 
\cite{Pullin}. (This result contains a source term that remains
inconsistent with our results.) With our general method it would take relatively little effort to examine this for $any$ second order 
multipole\footnote{Gleiser, \cite{GL2}, for example, has considered the linear quadrapole in conjunction with the second order 
monopole perturbation. This is essentially an investigation of how the second order mass perturbation is affected by the first order 
gravitational radiation.}, or even to reproduce the general result from \cite{Pullin}. Our methods, of course, are not limited to 
examining perturbations that have only a linear quadrapole contribution. Indeed, given the computational speed with which our simple 
``pure quadrapole'' case was solved\footnote{The calculations presented in
Appendix B were performed within a Maple 
worksheet running on a 600 MHz Alpha. The total CPU time was 
just under two minutes.}, there is no reason to believe that more 
realistic perturbations could not be examined. One could, for example, study how the inclusion of the first few multipoles at the linear 
level affect any given second order multipole. One might even consider higher
order calculations.
Given the minimal computational requirements of our second order 
calculations it is likely that third order Schwarzschild perturbations would
be manageable. While the utility of a third order calculation is certainly
questionable, the potential for its investigation does exist.

The largest obstacle to such calculations, and even to our present calculation, is finding the correct strategy for eliminating the 
redundancy in the linear perturbation equations. With some forethought, however, one could create a Maple routine that 
would examine a system of redundant PDEs, such as the ones we have encountered, and automatically generate a list of 
simplifying equations. One of these expressions might, for example, allow one to eliminate a second order derivative in terms of a linear 
combination of first order derivatives. Such a routine would certainly go a long way towards making these kinds of calculations much 
easier. 

With the methodology of Schwarzschild perturbations on such firm ground one is
quite naturally led to carrying this analysis over to the calculation of 
perturbations to the Kerr metric. As mentioned in the introduction, this
approach fails almost immediately when applied to the Kerr spacetime. The 
reason for this is simple: The background Kerr metric is not spherically 
symmetric, which results in perturbation equations that are not separable 
and that can therefore not be decoupled. 
One is free to try and find a coordinate system in
which the equations become separable, but one should be forewarned that 
this is not a
rewarding pursuit. None of the common coordinate systems of the Kerr spacetime
produce separable perturbation equations. Unless one can guess, or derive, such
a coordinate system, this approach to the Kerr problem is a dead end. It thus
appears that one must appeal to the tetrad solution of the linear Kerr problem
for guidance.

The central result of the tetrad analysis is the Teukolsky equation 
(see, $e.g.$, \cite{T}), 
${\cal L}_T \psi_i = 0$, where ${\cal L}_T$ is a complex linear operator and 
the $\psi_i$ are the $\psi_0$ and $\psi_4/\rho^4$ scalars from the NP 
formalism. All of the perturbation information is therefore encoded in just two
complex scalars, so that one expects that there are only four independent MP
functions. Our approach to this problem is to try and reverse engineer the NP
solution and express the metric perturbations in terms of the perturbed NP 
quantities. If this can be done we will be assured that the resulting 
linear perturbation equations will decouple. By examining how the combinations
of the separable eigenfunctions appear in the resulting linearized Ricci tensor
one could then develop an algorithm, similar to our Schwarzschild result, to 
decouple the linear perturbations of the Kerr spacetime. If this linear problem
can be solved with ``reasonable'' computing resources then one should be able 
to repeat the analysis for second order Kerr perturbations, just as we have 
done for the Schwarzschild case.

Our preliminary calculations in this effort indicate that even the linear Kerr
problem can generate extremely large and complex expressions. Although much of
this complexity disappears after extensive simplification. Despite the size of
some of these intermediate objects, we are confident that the linear Kerr 
problem can be solved by this ``reverse engineering'' method. 
Our work is now proceeding in this direction, and we expect to have definite 
Kerr 
results shortly. Our findings here will be the subject of Paper II. 
If the linear Kerr 
calculation requires a large percentage of our full computing resources then
it is likely that our findings for the second order Kerr case will simply be 
that a full solution to the problem will have to wait for 
the next generation of computer/computing engine. 
As our final comment we wish to
note that the difficulty and complexity of the linear Kerr problem can be 
truly surprising. 

\subsubsection*{Acknowledgments}

The results presented here have grown out of work that I have done for an
advanced graduate course in General Relativity given by Kayll Lake. It was
Kayll who first peaked my interest in this problem, and I greatfully 
acknowledge many useful conversations with both Kayll and Nicos Pelavas. I also
thank Jorge Pullin and Reinaldo J. Gleiser for reading the manuscript.

\newpage

\subsection*{Appendix A}

This appendix contains some of the details of our new Maple routines that were used in our calculations. Most of these routines are fairly 
small, and did not require a significant amount of coding. 
(There are, however, some notable exceptions.) 
Our examples\footnote{
While these examples may seem 
somewhat artificial to the uninitiated, one should bear in mind that the result
of a large algebraic calculation is typically an almost fully expanded 
expression that must be simplified. 
} will make use of the 
following two polynomials that are similar to those found in our calculations:

\bigskip

\begin{mapleinput}
p1:=(r^2+M*r+M^2)^2*(r-2*M)^2/r^2/(2*r+3*M)^2;
\end{mapleinput}
\begin{maplelatex}
\[
{\it p1} := {\displaystyle \frac {(\,{r}^{2} + {M}\,{r} + {M}^{2}
\,)^{2}\,(\,{r} - 2\,{M}\,)^{2}}{{r}^{2}\,(\,2\,{r} + 3\,{M}\,)^{
2}}}
\]
\end{maplelatex}
\begin{mapleinput}
p2:=(2*r^2-6*M*r-3*M^2)^2*(r-2*M)^2/r/(r-2*M)^2/(2*r+3*M);
\end{mapleinput}
\begin{maplelatex}
\[
{\it p2} := {\displaystyle \frac {(\,2\,{r}^{2} - 6\,{M}\,{r} - 3
\,{M}^{2}\,)^{2}}{{r}\,(\,2\,{r} + 3\,{M}\,)}}
\]
\end{maplelatex}

\noindent $\bullet$ {\bf termsimp({\em expr})}: The {\bf termsimp()} routine applies {\bf factor(simplify())} to each term in a type {\bf $+$} 
or {\bf $*$} expression individually, and then reconstructs the final expression. This can be very useful if one needs to simplify an expression 
after having applied the {\bf collect()} routine, as Maple's {\bf simplify()} and {\bf collect()} tend to be inverse operations.

{\em e.g.}

\begin{mapleinput}
temp:=expand(p1)*H[0]+expand(p2)*H[1];
\end{mapleinput}
\begin{maplelatex}
\begin{eqnarray*}
\lefteqn{{\it temp} :=  \left( {\vrule 
height0.87em width0em depth0.87em} \right. \! \! {\displaystyle 
\frac {{r}^{4}}{(\,2\,{r} + 3\,{M}\,)^{2}}} - 2\,{\displaystyle 
\frac {{r}^{3}\,{M}}{(\,2\,{r} + 3\,{M}\,)^{2}}} - 
{\displaystyle \frac {{r}^{2}\,{M}^{2}}{(\,2\,{r} + 3\,{M}\,)^{2}
}} - 2\,{\displaystyle \frac {{r}\,{M}^{3}}{(\,2\,{r} + 3\,{M}\,)
^{2}}} + 5\,{\displaystyle \frac {{M}^{4}}{(\,2\,{r} + 3\,{M}\,)
^{2}}}} \\
 & & \mbox{} + 4\,{\displaystyle \frac {{M}^{5}}{{r}\,(\,2\,{r}
 + 3\,{M}\,)^{2}}} + 4\,{\displaystyle \frac {{M}^{6}}{{r}^{2}\,(
\,2\,{r} + 3\,{M}\,)^{2}}} \! \! \left. {\vrule 
height0.87em width0em depth0.87em} \right) {{H}_{0}} \\
 & & \mbox{} +  \left( \! \,4\,{\displaystyle \frac {{r}^{3}}{2\,
{r} + 3\,{M}}} - 24\,{\displaystyle \frac {{r}^{2}\,{M}}{2\,{r}
 + 3\,{M}}} + 24\,{\displaystyle \frac {{r}\,{M}^{2}}{2\,{r} + 3
\,{M}}} + 36\,{\displaystyle \frac {{M}^{3}}{2\,{r} + 3\,{M}}} + 
9\,{\displaystyle \frac {{M}^{4}}{{r}\,(\,2\,{r} + 3\,{M}\,)}}\,
 \!  \right) \,{{H}_{1}}
\end{eqnarray*}
\end{maplelatex}
\begin{mapleinput}
termsimp(temp);
\end{mapleinput}
\begin{maplelatex}
\[
{\displaystyle \frac {(\,{r} - 2\,{M}\,)^{2}\,(\,{r}^{2} + {M}\,{
r} + {M}^{2}\,)^{2}\,{{H}_{0}}}{{r}^{2}\,(\,2\,{r} + 3\,{M}\,)^{2
}}} + {\displaystyle \frac {(\,2\,{r}^{2} - 6\,{M}\,{r} - 3\,{M}
^{2}\,)^{2}\,{{H}_{1}}}{{r}\,(\,2\,{r} + 3\,{M}\,)}}
\]
\end{maplelatex}

\bigskip

\noindent We can now compare this with the result of applying the Maple {\bf simplify()} routine.

\bigskip

\begin{mapleinput}
simplify(temp);
\end{mapleinput}
\begin{maplelatex}
\begin{eqnarray*}
\lefteqn{ \left( {\vrule height0.47em width0em depth0.47em}
 \right. \! \! {{H}_{0}}\,{r}^{6} - 2\,{{H}_{0}}\,{r}^{5}\,{M} - 
{{H}_{0}}\,{r}^{4}\,{M}^{2} - 2\,{{H}_{0}}\,{r}^{3}\,{M}^{3} + 5
\,{{H}_{0}}\,{r}^{2}\,{M}^{4} + 4\,{{H}_{0}}\,{M}^{5}\,{r} + 4\,{
{H}_{0}}\,{M}^{6} + 8\,{{H}_{1}}\,{r}^{6}} \\
 & & \mbox{} - 36\,{{H}_{1}}\,{r}^{5}\,{M} - 24\,{{H}_{1}}\,{r}^{
4}\,{M}^{2} + 144\,{{H}_{1}}\,{r}^{3}\,{M}^{3} + 126\,{{H}_{1}}\,
{r}^{2}\,{M}^{4} + 27\,{{H}_{1}}\,{r}\,{M}^{5} \! \! \left. 
{\vrule height0.47em width0em depth0.47em} \right)  \left/ 
{\vrule height0.44em width0em depth0.44em} \right. \! \!  \left( 
{\vrule height0.44em width0em depth0.44em} \right. \! \! \,{r}^{2
} \\
 & & (\,2\,{r} + 3\,{M}\,)^{2}\, \! \! \left. {\vrule 
height0.44em width0em depth0.44em} \right) \mbox{\hspace{328pt}}
\end{eqnarray*}
\end{maplelatex}

\noindent This rather cumbersome expression is a result of the fact that {\bf simplify()} does not ``know'' that the $H_i$'s should be treated as the 
primary objects. The effect of using {\bf collect()} (which would produce an 
expression similar to {\bf temp}) and then {\bf termsimp()} essentially tells
Maple that the ``collected'' objects are the primary ones.

\bigskip

\noindent $\bullet$ {\bf hcollect({\em expr, list\_one, list\_two})}: The {\bf hcollect()} routine is a hierarchal collection routine. The {\bf 
collect()} routine is first applied to the expression with {\em list\_one} as an argument. The resulting expression, which has been factored 
w.r.t. the elements of {\em list\_one}, is then collected w.r.t. the elements of {\em list\_two}. The utility of this function is that it is designed 
not to undo any of the first collection, which is the typical result of 
applying successive {\bf collect()} calls with different lists.
This routine automatically calls {\bf termsimp()} when reconstructing the
final expression.

{\em e.g.}

\begin{mapleinput}
temp:=expand(p1*(u(r)+(r-2*M)/r*diff(u(r),r))*H[0]+
p2*(u(r)+(2*r+3*M)*diff(u(r),r))*H[1]);
\end{mapleinput}
\begin{maplelatex}
\begin{eqnarray*}
\lefteqn{{\it temp} := {\displaystyle \frac {{r}^{4}\,{{H}_{0}}\,
{\rm u}(\,{r}\,)}{(\,2\,{r} + 3\,{M}\,)^{2}}} + {\displaystyle 
\frac {{r}^{4}\,{{H}_{0}}\, \left( \! \,{\frac {{ \partial}}{{ 
\partial}{r}}}\,{\rm u}(\,{r}\,)\, \!  \right) }{(\,2\,{r} + 3\,{
M}\,)^{2}}} - 4\,{\displaystyle \frac {{r}^{3}\,{{H}_{0}}\,
 \left( \! \,{\frac {{ \partial}}{{ \partial}{r}}}\,{\rm u}(\,{r}
\,)\, \!  \right) \,{M}}{(\,2\,{r} + 3\,{M}\,)^{2}}} - 2\,
{\displaystyle \frac {{r}^{3}\,{{H}_{0}}\,{M}\,{\rm u}(\,{r}\,)}{
(\,2\,{r} + 3\,{M}\,)^{2}}}} \\
 & & \mbox{} + 3\,{\displaystyle \frac {{r}^{2}\,{{H}_{0}}\,{M}^{
2}\, \left( \! \,{\frac {{ \partial}}{{ \partial}{r}}}\,{\rm u}(
\,{r}\,)\, \!  \right) }{(\,2\,{r} + 3\,{M}\,)^{2}}} - 
{\displaystyle \frac {{r}^{2}\,{{H}_{0}}\,{M}^{2}\,{\rm u}(\,{r}
\,)}{(\,2\,{r} + 3\,{M}\,)^{2}}} - 2\,{\displaystyle \frac {{r}\,
{{H}_{0}}\,{M}^{3}\,{\rm u}(\,{r}\,)}{(\,2\,{r} + 3\,{M}\,)^{2}}}
 + 9\,{\displaystyle \frac {{{H}_{0}}\,{M}^{4}\, \left( \! \,
{\frac {{ \partial}}{{ \partial}{r}}}\,{\rm u}(\,{r}\,)\, \! 
 \right) }{(\,2\,{r} + 3\,{M}\,)^{2}}} \\
 & & \mbox{} + 5\,{\displaystyle \frac {{{H}_{0}}\,{M}^{4}\,{\rm 
u}(\,{r}\,)}{(\,2\,{r} + 3\,{M}\,)^{2}}} - 6\,{\displaystyle 
\frac {{{H}_{0}}\,{M}^{5}\, \left( \! \,{\frac {{ \partial}}{{ 
\partial}{r}}}\,{\rm u}(\,{r}\,)\, \!  \right) }{{r}\,(\,2\,{r}
 + 3\,{M}\,)^{2}}} + 4\,{\displaystyle \frac {{{H}_{0}}\,{M}^{5}
\,{\rm u}(\,{r}\,)}{{r}\,(\,2\,{r} + 3\,{M}\,)^{2}}} - 4\,
{\displaystyle \frac {{{H}_{0}}\,{M}^{6}\, \left( \! \,{\frac {{ 
\partial}}{{ \partial}{r}}}\,{\rm u}(\,{r}\,)\, \!  \right) }{{r}
^{2}\,(\,2\,{r} + 3\,{M}\,)^{2}}} \\
 & & \mbox{} + 4\,{\displaystyle \frac {{{H}_{0}}\,{M}^{6}\,{\rm 
u}(\,{r}\,)}{{r}^{2}\,(\,2\,{r} + 3\,{M}\,)^{2}}} - 8\,
{\displaystyle \frac {{{H}_{0}}\,{M}^{7}\, \left( \! \,{\frac {{ 
\partial}}{{ \partial}{r}}}\,{\rm u}(\,{r}\,)\, \!  \right) }{{r}
^{3}\,(\,2\,{r} + 3\,{M}\,)^{2}}} + 4\,{\displaystyle \frac {{r}
^{3}\,{{H}_{1}}\,{\rm u}(\,{r}\,)}{2\,{r} + 3\,{M}}} + 8\,
{\displaystyle \frac {{r}^{4}\,{{H}_{1}}\, \left( \! \,{\frac {{ 
\partial}}{{ \partial}{r}}}\,{\rm u}(\,{r}\,)\, \!  \right) }{2\,
{r} + 3\,{M}}} \\
 & & \mbox{} - 36\,{\displaystyle \frac {{r}^{3}\,{{H}_{1}}\,
 \left( \! \,{\frac {{ \partial}}{{ \partial}{r}}}\,{\rm u}(\,{r}
\,)\, \!  \right) \,{M}}{2\,{r} + 3\,{M}}} - 24\,{\displaystyle 
\frac {{r}^{2}\,{{H}_{1}}\,{M}\,{\rm u}(\,{r}\,)}{2\,{r} + 3\,{M}
}} - 24\,{\displaystyle \frac {{r}^{2}\,{{H}_{1}}\,{M}^{2}\,
 \left( \! \,{\frac {{ \partial}}{{ \partial}{r}}}\,{\rm u}(\,{r}
\,)\, \!  \right) }{2\,{r} + 3\,{M}}} + 24\,{\displaystyle 
\frac {{r}\,{{H}_{1}}\,{M}^{2}\,{\rm u}(\,{r}\,)}{2\,{r} + 3\,{M}
}} \\
 & & \mbox{} + 144\,{\displaystyle \frac {{r}\,{{H}_{1}}\,{M}^{3}
\, \left( \! \,{\frac {{ \partial}}{{ \partial}{r}}}\,{\rm u}(\,{
r}\,)\, \!  \right) }{2\,{r} + 3\,{M}}} + 36\,{\displaystyle 
\frac {{{H}_{1}}\,{M}^{3}\,{\rm u}(\,{r}\,)}{2\,{r} + 3\,{M}}} + 
126\,{\displaystyle \frac {{{H}_{1}}\,{M}^{4}\, \left( \! \,
{\frac {{ \partial}}{{ \partial}{r}}}\,{\rm u}(\,{r}\,)\, \! 
 \right) }{2\,{r} + 3\,{M}}} + 9\,{\displaystyle \frac {{{H}_{1}}
\,{M}^{4}\,{\rm u}(\,{r}\,)}{{r}\,(\,2\,{r} + 3\,{M}\,)}} \\
 & & \mbox{} + 27\,{\displaystyle \frac {{{H}_{1}}\,{M}^{5}\,
 \left( \! \,{\frac {{ \partial}}{{ \partial}{r}}}\,{\rm u}(\,{r}
\,)\, \!  \right) }{{r}\,(\,2\,{r} + 3\,{M}\,)}}
\end{eqnarray*}
\end{maplelatex}
\begin{mapleinput}
hcollect(temp,{u(r),diff(u(r),r)},{H[0],H[1]});
\end{mapleinput}
\begin{maplelatex}
\begin{eqnarray*}
\lefteqn{ \left( \! \,{\displaystyle \frac {(\,{r}^{2} + {M}\,{r}
 + {M}^{2}\,)^{2}\,(\,{r} - 2\,{M}\,)^{3}\,{{H}_{0}}}{{r}^{3}\,(
\,2\,{r} + 3\,{M}\,)^{2}}} + {\displaystyle \frac {(\,2\,{r}^{2}
 - 6\,{M}\,{r} - 3\,{M}^{2}\,)^{2}\,{{H}_{1}}}{{r}}}\, \! 
 \right) \, \left( \! \,{\frac {{ \partial}}{{ \partial}{r}}}\,
{\rm u}(\,{r}\,)\, \!  \right) } \\
 & & \mbox{} +  \left( \! \,{\displaystyle \frac {(\,{r} - 2\,{M}
\,)^{2}\,(\,{r}^{2} + {M}\,{r} + {M}^{2}\,)^{2}\,{{H}_{0}}}{{r}^{
2}\,(\,2\,{r} + 3\,{M}\,)^{2}}} + {\displaystyle \frac {(\,2\,{r}
^{2} - 6\,{M}\,{r} - 3\,{M}^{2}\,)^{2}\,{{H}_{1}}}{{r}\,(\,2\,{r}
 + 3\,{M}\,)}}\, \!  \right) \,{\rm u}(\,{r}\,)
\end{eqnarray*}
\end{maplelatex}

\noindent As with any ``free code'', one should always perform some random
consistency checks:

\bigskip

\begin{mapleinput}
simplify(hcollect(temp,{u(r),diff(u(r),r)},{H[0],H[1]})-temp);
\end{mapleinput}
\begin{maplelatex}
\[
0
\]
\end{maplelatex}

\bigskip

\noindent $\bullet$ {\bf mcollect({\em expr}, \{{\em arg\_one} [, {\em arg\_two} ]\})}: The {\bf mcollect()} routine will collect the terms of 
an expression w.r.t. the product $(${\em arg\_one arg\_two}$)$. If {\em arg\_two} is not supplied the collection is done w.r.t. $(${\em 
arg\_one}$)^2$. (This routine required a relatively large amount of coding. In particular, one has to very careful, from a coding standpoint, 
when the arguments can be derivatives of a function.) 
This routine automatically calls {\bf termsimp()} when reconstructing the
final expression.

{\em e.g.}

\begin{mapleinput}
temp:=expand(p1*(u(r)+(r-2*M)/r*diff(u(r),r))*(u(r)+(r-2*M)/r*diff(u(r),r)));
\end{mapleinput}
\begin{maplelatex}
\begin{eqnarray*}
\lefteqn{{\it temp} := 2\,{\displaystyle \frac {{r}^{4}\,{\rm u}(
\,{r}\,)\, \left( \! \,{\frac {{ \partial}}{{ \partial}{r}}}\,
{\rm u}(\,{r}\,)\, \!  \right) }{(\,2\,{r} + 3\,{M}\,)^{2}}} + 11
\,{\displaystyle \frac {{r}^{2}\, \left( \! \,{\frac {{ \partial}
}{{ \partial}{r}}}\,{\rm u}(\,{r}\,)\, \!  \right) ^{2}\,{M}^{2}
}{(\,2\,{r} + 3\,{M}\,)^{2}}} - 2\,{\displaystyle \frac {{r}^{3}
\,{M}\,{\rm u}(\,{r}\,)^{2}}{(\,2\,{r} + 3\,{M}\,)^{2}}} - 6\,
{\displaystyle \frac {{r}\,{M}^{3}\, \left( \! \,{\frac {{ 
\partial}}{{ \partial}{r}}}\,{\rm u}(\,{r}\,)\, \!  \right) ^{2}
}{(\,2\,{r} + 3\,{M}\,)^{2}}}} \\
 & & \mbox{} - {\displaystyle \frac {{r}^{2}\,{M}^{2}\,{\rm u}(\,
{r}\,)^{2}}{(\,2\,{r} + 3\,{M}\,)^{2}}} + 9\,{\displaystyle 
\frac {{M}^{4}\, \left( \! \,{\frac {{ \partial}}{{ \partial}{r}
}}\,{\rm u}(\,{r}\,)\, \!  \right) ^{2}}{(\,2\,{r} + 3\,{M}\,)^{2
}}} - 2\,{\displaystyle \frac {{r}\,{M}^{3}\,{\rm u}(\,{r}\,)^{2}
}{(\,2\,{r} + 3\,{M}\,)^{2}}} - 24\,{\displaystyle \frac {{M}^{5}
\, \left( \! \,{\frac {{ \partial}}{{ \partial}{r}}}\,{\rm u}(\,{
r}\,)\, \!  \right) ^{2}}{{r}\,(\,2\,{r} + 3\,{M}\,)^{2}}} \\
 & & \mbox{} + 5\,{\displaystyle \frac {{M}^{4}\,{\rm u}(\,{r}\,)
^{2}}{(\,2\,{r} + 3\,{M}\,)^{2}}} + 8\,{\displaystyle \frac {{M}
^{6}\, \left( \! \,{\frac {{ \partial}}{{ \partial}{r}}}\,{\rm u}
(\,{r}\,)\, \!  \right) ^{2}}{{r}^{2}\,(\,2\,{r} + 3\,{M}\,)^{2}
}} + 4\,{\displaystyle \frac {{M}^{5}\,{\rm u}(\,{r}\,)^{2}}{{r}
\,(\,2\,{r} + 3\,{M}\,)^{2}}} + 16\,{\displaystyle \frac {{M}^{8}
\, \left( \! \,{\frac {{ \partial}}{{ \partial}{r}}}\,{\rm u}(\,{
r}\,)\, \!  \right) ^{2}}{{r}^{4}\,(\,2\,{r} + 3\,{M}\,)^{2}}} \\
 & & \mbox{} + {\displaystyle \frac {{r}^{4}\,{\rm u}(\,{r}\,)^{2
}}{(\,2\,{r} + 3\,{M}\,)^{2}}} + {\displaystyle \frac {{r}^{4}\,
 \left( \! \,{\frac {{ \partial}}{{ \partial}{r}}}\,{\rm u}(\,{r}
\,)\, \!  \right) ^{2}}{(\,2\,{r} + 3\,{M}\,)^{2}}} + 4\,
{\displaystyle \frac {{M}^{6}\,{\rm u}(\,{r}\,)^{2}}{{r}^{2}\,(\,
2\,{r} + 3\,{M}\,)^{2}}} - 8\,{\displaystyle \frac {{r}^{3}\,
{\rm u}(\,{r}\,)\, \left( \! \,{\frac {{ \partial}}{{ \partial}{r
}}}\,{\rm u}(\,{r}\,)\, \!  \right) \,{M}}{(\,2\,{r} + 3\,{M}\,)
^{2}}} \\
 & & \mbox{} - 6\,{\displaystyle \frac {{r}^{3}\, \left( \! \,
{\frac {{ \partial}}{{ \partial}{r}}}\,{\rm u}(\,{r}\,)\, \! 
 \right) ^{2}\,{M}}{(\,2\,{r} + 3\,{M}\,)^{2}}} + 6\,
{\displaystyle \frac {{r}^{2}\,{M}^{2}\,{\rm u}(\,{r}\,)\,
 \left( \! \,{\frac {{ \partial}}{{ \partial}{r}}}\,{\rm u}(\,{r}
\,)\, \!  \right) }{(\,2\,{r} + 3\,{M}\,)^{2}}} + 18\,
{\displaystyle \frac {{M}^{4}\,{\rm u}(\,{r}\,)\, \left( \! \,
{\frac {{ \partial}}{{ \partial}{r}}}\,{\rm u}(\,{r}\,)\, \! 
 \right) }{(\,2\,{r} + 3\,{M}\,)^{2}}} \\
 & & \mbox{} - 12\,{\displaystyle \frac {{M}^{5}\,{\rm u}(\,{r}\,
)\, \left( \! \,{\frac {{ \partial}}{{ \partial}{r}}}\,{\rm u}(\,
{r}\,)\, \!  \right) }{{r}\,(\,2\,{r} + 3\,{M}\,)^{2}}} - 8\,
{\displaystyle \frac {{M}^{6}\,{\rm u}(\,{r}\,)\, \left( \! \,
{\frac {{ \partial}}{{ \partial}{r}}}\,{\rm u}(\,{r}\,)\, \! 
 \right) }{{r}^{2}\,(\,2\,{r} + 3\,{M}\,)^{2}}} - 16\,
{\displaystyle \frac {{M}^{7}\,{\rm u}(\,{r}\,)\, \left( \! \,
{\frac {{ \partial}}{{ \partial}{r}}}\,{\rm u}(\,{r}\,)\, \! 
 \right) }{{r}^{3}\,(\,2\,{r} + 3\,{M}\,)^{2}}}
\mbox{\hspace{48pt}}
\end{eqnarray*}
\end{maplelatex}
\begin{mapleinput}
mtemp:=mcollect(mcollect(mcollect(
mcollect(temp,{diff(u(r),r)}),{diff(u(r),r),u(r)}),{u(r)}),{diff(u(r),r)});
\end{mapleinput}
\begin{maplelatex}
\begin{eqnarray*}
\lefteqn{{\it mtemp} := {\displaystyle \frac {(\,{r}^{2} + {M}\,{
r} + {M}^{2}\,)^{2}\,(\,{r} - 2\,{M}\,)^{2}\,{\rm u}(\,{r}\,)^{2}
}{{r}^{2}\,(\,2\,{r} + 3\,{M}\,)^{2}}} + 2\,{\displaystyle 
\frac {(\,{r}^{2} + {M}\,{r} + {M}^{2}\,)^{2}\,(\,{r} - 2\,{M}\,)
^{3}\,{\rm u}(\,{r}\,)\, \left( \! \,{\frac {{ \partial}}{{ 
\partial}{r}}}\,{\rm u}(\,{r}\,)\, \!  \right) }{{r}^{3}\,(\,2\,{
r} + 3\,{M}\,)^{2}}}} \\
 & & \mbox{} + {\displaystyle \frac {(\,{r}^{2} + {M}\,{r} + {M}
^{2}\,)^{2}\,(\,{r} - 2\,{M}\,)^{4}\, \left( \! \,{\frac {{ 
\partial}}{{ \partial}{r}}}\,{\rm u}(\,{r}\,)\, \!  \right) ^{2}
}{{r}^{4}\,(\,2\,{r} + 3\,{M}\,)^{2}}}\mbox{\hspace{214pt}}
\end{eqnarray*}
\end{maplelatex}

\noindent And just to be sure:

\bigskip

\begin{mapleinput}
simplify(mtemp-temp);
\end{mapleinput}
\begin{maplelatex}
\[
0
\]
\end{maplelatex}

\bigskip

\noindent $\bullet$ {\bf kfactor({\em expr, fctr})}: The {\bf kfactor()} routine simply forces Maple to pull the factor {\em fctr} out of the 
expression. The chief utility of this is largely cosmetic, although it can be useful when comparing expressions to published results.

{\em e.g.}

\begin{mapleinput}
temp:=p1*H[0]+p2*H[1];
\end{mapleinput}
\begin{maplelatex}
\[
{\it temp} := {\displaystyle \frac {(\,{r}^{2} + {M}\,{r} + {M}^{
2}\,)^{2}\,(\,{r} - 2\,{M}\,)^{2}\,{{H}_{0}}}{{r}^{2}\,(\,2\,{r}
 + 3\,{M}\,)^{2}}} + {\displaystyle \frac {(\,2\,{r}^{2} - 6\,{M}
\,{r} - 3\,{M}^{2}\,)^{2}\,{{H}_{1}}}{{r}\,(\,2\,{r} + 3\,{M}\,)
}}
\]
\end{maplelatex}
\begin{mapleinput}
kfactor(temp,(r^2+M*r+M^2)^2*(r-2*M)^2);
\end{mapleinput}
\begin{maplelatex}
\[
(\,{r}^{2} + {M}\,{r} + {M}^{2}\,)^{2}\,(\,{r} - 2\,{M}\,)^{2}\,
 \left( \! \,{\displaystyle \frac {{{H}_{0}}}{{r}^{2}\,(\,2\,{r}
 + 3\,{M}\,)^{2}}} + {\displaystyle \frac {(\,2\,{r}^{2} - 6\,{M}
\,{r} - 3\,{M}^{2}\,)^{2}\,{{H}_{1}}}{{r}\,(\,2\,{r} + 3\,{M}\,)
\,(\,{r}^{2} + {M}\,{r} + {M}^{2}\,)^{2}\,(\,{r} - 2\,{M}\,)^{2}
}}\, \!  \right) 
\]
\end{maplelatex}

\bigskip

\noindent $\bullet$ {\bf linpert({\em expr})}: The {\bf linpert()} routine returns the first order Taylor expansion in $\epsilon$, about 
$\epsilon=0$, of the supplied expression. While this routine could be made far more elaborate in terms of options and parameters, it was 
specifically designed as a single argument function so that it could be used within GRTensorII in a seamless manner. 

{\em e.g.}

\begin{mapleinput}
temp:=1/sqrt(1+epsilon*f(r,theta,phi,t));
\end{mapleinput}
\begin{maplelatex}
\[
{\it temp} := {\displaystyle \frac {1}{\sqrt {1 + { \varepsilon}
\,{\rm f}(\,{r}, { \theta}, { \phi}, {t}\,)}}}
\]
\end{maplelatex}
\begin{mapleinput}
linpert(temp);
\end{mapleinput}
\begin{maplelatex}
\[
1 - {\displaystyle \frac {1}{2}}\,{ \varepsilon}\,{\rm f}(\,{r}, 
{ \theta}, { \phi}, {t}\,)
\]
\end{maplelatex}

\bigskip

\noindent $\bullet$ {\bf quadpert({\em expr})}: The {\bf quadpert()} routine is similar to the {\bf linpert()} routine, except that it returns the 
second order Taylor expansion of the expression in $\epsilon$.

{\em e.g.}

\begin{mapleinput}
quadpert(temp);
\end{mapleinput}
\begin{maplelatex}
\[
1 - {\displaystyle \frac {1}{2}}\,{ \varepsilon}\,{\rm f}(\,{r}, 
{ \theta}, { \phi}, {t}\,) + {\displaystyle \frac {3}{8}}\,{\rm f
}(\,{r}, { \theta}, { \phi}, {t}\,)^{2}\,{ \varepsilon}^{2}
\]
\end{maplelatex}

\end{document}


\setcounter{page}{21}
\subsection*{Appendix B}
\bigskip
The bulk of this appendix is the output from a Maple worksheet that contains
many of the ``gritty'' details of our second order Schwarzschild calculation.
A Maple worksheet that reproduces the entire calculation is available for
download at the GRTensor website.
\bigskip
\begin{maplegroup}
\begin{mapleinput}
\mapleinline{active}{1d}{restart:}{%
}
\end{mapleinput}

\end{maplegroup}
\begin{maplegroup}

\begin{mapleinput}
\mapleinline{active}{1d}{grtw():}{%
}
\end{mapleinput}

\mapleresult
\begin{maplelatex}
\[
\mathit{GRTensorII\ Version\ 1.70\ (R5)}
\]
\end{maplelatex}

\begin{maplelatex}
\[
\mathit{31\ May\ 1998}
\]
\end{maplelatex}

\begin{maplelatex}
\[
\mathit{Developed\ by\ Peter\ Musgrave,\ Denis\ Pollney\ and\ 
Kayll\ Lake}
\]
\end{maplelatex}

\begin{maplelatex}
\[
\mathit{Copyright\ 1994-1998\ by\ the\ authors.}
\]
\end{maplelatex}

\begin{maplelatex}
\[
\mathit{Latest\ version\ available\ from:\ 
http://astro.queensu.ca/\symbol{126}grtensor/}
\]
\end{maplelatex}

\end{maplegroup}
\begin{maplegroup}
\begin{mapleinput}
\mapleinline{active}{1d}{mine():}{%
}
\end{mapleinput}

\end{maplegroup}
\begin{maplegroup}
\begin{mapleinput}
\mapleinline{active}{1d}{read `myutils.mpl`;}{%
}
\end{mapleinput}

\end{maplegroup}
\begin{maplegroup}

\begin{mapleinput}
\mapleinline{active}{1d}{qload(qschw):}{%
}
\end{mapleinput}

\mapleresult
\begin{maplelatex}
\[
\mathit{Default\ spacetime}=\mathit{qschw}
\]
\end{maplelatex}

\begin{maplelatex}
\[
\mathit{For\ the\ qschw\ spacetime:}
\]
\end{maplelatex}

\begin{maplelatex}
\[
\mathit{Coordinates}
\]
\end{maplelatex}

\begin{maplelatex}
\[
\mathrm{x}(\mathit{up})
\]
\end{maplelatex}

\begin{maplelatex}
\[
\mathit{x\ }^{a}=[r, \,\theta , \,\phi , \,t]
\]
\end{maplelatex}

\begin{maplelatex}
\[
\mathit{Line\ element}
\]
\end{maplelatex}

\begin{maplelatex}
\begin{eqnarray*}
\lefteqn{\mathit{\ ds}^{2}={\displaystyle \frac {(1 + 
{\displaystyle \frac {1}{2}} \,(\varepsilon \,{h_{2}}(r, \,t) + 
\varepsilon ^{2}\,{H_{2}}(r, \,t))\,\mathrm{\%1})\,\mathit{\ d}\,
r^{\mathit{2\ }}}{1 - 2\,{\displaystyle \frac {m}{r}} }}  + (
\varepsilon \,{h_{1}}(r, \,t) + \varepsilon ^{2}\,{H_{1}}(r, \,t)
)\,\mathrm{\%1}\,\mathit{\ d}\,r^{\ }\,\mathit{d\ }\,t^{\ }} \\
 & & \mbox{} + r^{2}\,(1 + {\displaystyle \frac {1}{2}} \,(
\varepsilon \,\mathrm{k}(r, \,t) + \varepsilon ^{2}\,\mathrm{K}(r
, \,t))\,\mathrm{\%1})\,\mathit{\ d}\,\theta ^{\mathit{2\ }} \\
 & & \mbox{} + r^{2}\,\mathrm{sin}(\theta )^{2}\,(1 + 
{\displaystyle \frac {1}{2}} \,(\varepsilon \,\mathrm{k}(r, \,t)
 + \varepsilon ^{2}\,\mathrm{K}(r, \,t))\,\mathrm{\%1})\,\mathit{
\ d}\,\phi ^{\mathit{2\ }} \\
 & & \mbox{} - (1 - 2\,{\displaystyle \frac {m}{r}} )\,(1 - 
{\displaystyle \frac {1}{2}} \,(\varepsilon \,{h_{0}}(r, \,t) + 
\varepsilon ^{2}\,{H_{0}}(r, \,t))\,\mathrm{\%1})\,\mathit{\ d}\,
t^{\mathit{2\ }} \\
 & & \mathrm{\%1} := 3\,\mathrm{cos}(\theta )^{2} - 1
\mbox{\hspace{308pt}}
\end{eqnarray*}
\end{maplelatex}

\begin{maplelatex}
\[
\mathit{Constraints}=[\varepsilon ^{3}=0, \,\varepsilon ^{4}=0, 
\,\varepsilon ^{5}=0, \,\varepsilon ^{6}=0, \,\varepsilon ^{7}=0
, \,\varepsilon ^{8}=0, \,\varepsilon ^{9}=0, \,\varepsilon ^{10}
=0]
\]
\end{maplelatex}

\end{maplegroup}
\begin{maplegroup}
\bigskip

\noindent As can be seen from the form of the metric, the lower case
functions are the first order perturbations, and the upper case
functions are the second order perturbations. To begin our analysis we
calculate the exact contravariant metric tensor, which is then
quadratically perturbed using the { \bf quadpert() }
routine.

\bigskip

\begin{mapleinput}
\mapleinline{active}{1d}{grcalc(g(up,up)); }{%
}
\end{mapleinput}

\mapleresult
\begin{maplelatex}
\[
\mathit{CPU\ Time\ }=.394
\]
\end{maplelatex}

\end{maplegroup}
\begin{maplegroup}
\begin{mapleinput}
\mapleinline{active}{1d}{gralter(g(up,up),quadpert,simplify,factor);}{
}
\end{mapleinput}

\mapleresult
\begin{maplettyout}
Component simplification of a GRTensorII object:
\end{maplettyout}

\begin{maplettyout}
 
\end{maplettyout}

\begin{maplettyout}
Applying routine quadpert to object g(up,up)
\end{maplettyout}

\begin{maplettyout}
Applying routine simplify to object g(up,up)
\end{maplettyout}

\begin{maplettyout}
Applying routine factor to object g(up,up)
\end{maplettyout}

\begin{maplelatex}
\[
\mathit{CPU\ Time\ }=.305
\]
\end{maplelatex}

\end{maplegroup}
\begin{maplegroup}
\begin{mapleinput}
\mapleinline{active}{1d}{grdisplay(g(up,up)); }{%
}
\end{mapleinput}

\mapleresult
\begin{maplelatex}
\[
\mathit{For\ the\ qschw\ spacetime:}
\]
\end{maplelatex}

\begin{maplelatex}
\[
\mathit{Contravariant\ metric\ tensor}
\]
\end{maplelatex}

\begin{maplelatex}
\[
\mathrm{g}(\mathit{up}, \,\mathit{up})
\]
\end{maplelatex}

\begin{maplelatex}
\begin{eqnarray*}
\lefteqn{\mathit{g\ }^{r}\,\ ^{r}={\displaystyle \frac {1}{4}} ((
 - r + 2\,m)( - 4 + 6\,{h_{2}}(r, \,t)\,\varepsilon \,\mathrm{cos
}(\theta )^{2} - 2\,\varepsilon \,{h_{2}}(r, \,t) - \varepsilon 
^{2}\,{h_{2}}(r, \,t)^{2} + 6\,\varepsilon ^{2}\,{h_{2}}(r, \,t)
^{2}\,\mathrm{cos}(\theta )^{2}} \\
 & & \mbox{} - 9\,\varepsilon ^{2}\,{h_{2}}(r, \,t)^{2}\,\mathrm{
cos}(\theta )^{4} - 6\,\varepsilon ^{2}\,{h_{1}}(r, \,t)^{2}\,
\mathrm{cos}(\theta )^{2} + 9\,\varepsilon ^{2}\,{h_{1}}(r, \,t)
^{2}\,\mathrm{cos}(\theta )^{4} + 6\,\varepsilon ^{2}\,{H_{2}}(r
, \,t)\,\mathrm{cos}(\theta )^{2} \\
 & & \mbox{} - 2\,\varepsilon ^{2}\,{H_{2}}(r, \,t) + \varepsilon
 ^{2}\,{h_{1}}(r, \,t)^{2}))/r
\end{eqnarray*}
\end{maplelatex}

\begin{maplelatex}
\begin{eqnarray*}
\lefteqn{\mathit{g\ }^{r}\,\ ^{t}= - {\displaystyle \frac {1}{4}
} \varepsilon \,(3\,\mathrm{cos}(\theta )^{2} - 1)(3\,\varepsilon
 \,{h_{1}}(r, \,t)\,{h_{2}}(r, \,t)\,\mathrm{cos}(\theta )^{2} - 
3\,\varepsilon \,{h_{1}}(r, \,t)\,{h_{0}}(r, \,t)\,\mathrm{cos}(
\theta )^{2} - 2\,\varepsilon \,{H_{1}}(r, \,t)} \\
 & & \mbox{} + \varepsilon \,{h_{1}}(r, \,t)\,{h_{0}}(r, \,t) - 
\varepsilon \,{h_{1}}(r, \,t)\,{h_{2}}(r, \,t) - 2\,{h_{1}}(r, \,
t))\mbox{\hspace{184pt}}
\end{eqnarray*}
\end{maplelatex}

\begin{maplelatex}
\begin{eqnarray*}
\lefteqn{\mathit{g\ }^{\theta }\,\ ^{\theta }= - {\displaystyle 
\frac {1}{4}} ( - 4 + 6\,\varepsilon \,\mathrm{k}(r, \,t)\,
\mathrm{cos}(\theta )^{2} - 2\,\varepsilon \,\mathrm{k}(r, \,t)
 - 9\,\varepsilon ^{2}\,\mathrm{cos}(\theta )^{4}\,\mathrm{k}(r, 
\,t)^{2} + 6\,\varepsilon ^{2}\,\mathrm{k}(r, \,t)^{2}\,\mathrm{
cos}(\theta )^{2}} \\
 & & \mbox{} + 6\,\varepsilon ^{2}\,\mathrm{K}(r, \,t)\,\mathrm{
cos}(\theta )^{2} - \varepsilon ^{2}\,\mathrm{k}(r, \,t)^{2} - 2
\,\varepsilon ^{2}\,\mathrm{K}(r, \,t))/r^{2}
\mbox{\hspace{152pt}}
\end{eqnarray*}
\end{maplelatex}

\begin{maplelatex}
\begin{eqnarray*}
\lefteqn{\mathit{g\ }^{\phi }\,\ ^{\phi }={\displaystyle \frac {1
}{4}} ( - 4 + 6\,\varepsilon \,\mathrm{k}(r, \,t)\,\mathrm{cos}(
\theta )^{2} - 2\,\varepsilon \,\mathrm{k}(r, \,t) - 9\,
\varepsilon ^{2}\,\mathrm{cos}(\theta )^{4}\,\mathrm{k}(r, \,t)^{
2} + 6\,\varepsilon ^{2}\,\mathrm{k}(r, \,t)^{2}\,\mathrm{cos}(
\theta )^{2}} \\
 & & \mbox{} + 6\,\varepsilon ^{2}\,\mathrm{K}(r, \,t)\,\mathrm{
cos}(\theta )^{2} - \varepsilon ^{2}\,\mathrm{k}(r, \,t)^{2} - 2
\,\varepsilon ^{2}\,\mathrm{K}(r, \,t))/(r^{2}\,( - 1 + \mathrm{
cos}(\theta ))\,(\mathrm{cos}(\theta ) + 1))\mbox{\hspace{11pt}}
\end{eqnarray*}
\end{maplelatex}

\begin{maplelatex}
\begin{eqnarray*}
\lefteqn{\mathit{g\ }^{t}\,\ ^{t}={\displaystyle \frac {1}{4}} r(
4 + 6\,\varepsilon \,{h_{0}}(r, \,t)\,\mathrm{cos}(\theta )^{2}
 - 2\,\varepsilon \,{h_{0}}(r, \,t) - 6\,\varepsilon ^{2}\,{h_{0}
}(r, \,t)^{2}\,\mathrm{cos}(\theta )^{2} + 9\,\varepsilon ^{2}\,
\mathrm{cos}(\theta )^{4}\,{h_{0}}(r, \,t)^{2}} \\
 & & \mbox{} + \varepsilon ^{2}\,{h_{0}}(r, \,t)^{2} + 6\,
\varepsilon ^{2}\,{H_{0}}(r, \,t)\,\mathrm{cos}(\theta )^{2} + 6
\,\varepsilon ^{2}\,{h_{1}}(r, \,t)^{2}\,\mathrm{cos}(\theta )^{2
} - 9\,\varepsilon ^{2}\,{h_{1}}(r, \,t)^{2}\,\mathrm{cos}(\theta
 )^{4} \\
 & & \mbox{} - \varepsilon ^{2}\,{h_{1}}(r, \,t)^{2} - 2\,
\varepsilon ^{2}\,{H_{0}}(r, \,t))/( - r + 2\,m)
\mbox{\hspace{211pt}}
\end{eqnarray*}
\end{maplelatex}

\end{maplegroup}
\begin{maplegroup}

\bigskip
\noindent With these quadratically perturbed contra/co-variant
components of the metric we can now calculate the perturbed Ricci
tensor to second order.

\bigskip

\begin{mapleinput}
\mapleinline{active}{1d}{grcalcalter(R(dn,dn),13); }{%
}
\end{mapleinput}

\mapleresult
\begin{maplettyout}
Simplification will be applied during calculation.
\end{maplettyout}

\begin{maplettyout}
 
\end{maplettyout}

\begin{maplettyout}
Applying routine `Apply constraints repeatedly` to object g(dn,dn,pdn)
\end{maplettyout}

\begin{maplettyout}
Applying routine `Apply constraints repeatedly` to object
Chr(dn,dn,dn)
\end{maplettyout}

\begin{maplettyout}
Applying routine `Apply constraints repeatedly` to object
Chr(dn,dn,up)
\end{maplettyout}

\begin{maplettyout}
Applying routine `Apply constraints repeatedly` to object R(dn,dn)
\end{maplettyout}

\begin{maplelatex}
\[
\mathit{CPU\ Time\ }=5.811
\]
\end{maplelatex}

\end{maplegroup}
\begin{maplegroup}
\begin{mapleinput}
\mapleinline{active}{1d}{gralter(R(dn,dn),expand);}{%
}
\end{mapleinput}

\mapleresult
\begin{maplettyout}
Component simplification of a GRTensorII object:
\end{maplettyout}

\begin{maplettyout}
 
\end{maplettyout}

\begin{maplettyout}
Applying routine expand to object R(dn,dn)
\end{maplettyout}

\begin{maplelatex}
\[
\mathit{CPU\ Time\ }=.211
\]
\end{maplelatex}

\end{maplegroup}
\begin{maplegroup}
\begin{mapleinput}
\mapleinline{active}{1d}{grdisplay(R(dn,dn)); }{%
}
\end{mapleinput}

\mapleresult
\begin{maplelatex}
\[
\mathit{For\ the\ qschw\ spacetime:}
\]
\end{maplelatex}

\begin{maplelatex}
\[
\mathit{Covariant\ Ricci}
\]
\end{maplelatex}

\begin{maplelatex}
\[
\mathrm{R}(\mathit{dn}, \,\mathit{dn})
\]
\end{maplelatex}

\begin{maplelatex}
\[
{\mathit{R\ }_{r}}\,{\ _{r}}=\mathit{\ 86717\ words.\ Exceeds\ 
grOptionDisplayLimit}
\]
\end{maplelatex}

\begin{maplelatex}
\[
{\mathit{R\ }_{r}}\,{\ _{\theta }}=\mathit{\ 28579\ words.\ 
Exceeds\ grOptionDisplayLimit}
\]
\end{maplelatex}

\begin{maplelatex}
\[
{\mathit{R\ }_{r}}\,{\ _{t}}=\mathit{\ 53004\ words.\ Exceeds\ 
grOptionDisplayLimit}
\]
\end{maplelatex}

\begin{maplelatex}
\[
{\mathit{R\ }_{\theta }}\,{\ _{\theta }}=\mathit{\ 87057\ words.
\ Exceeds\ grOptionDisplayLimit}
\]
\end{maplelatex}

\begin{maplelatex}
\[
{\mathit{R\ }_{\theta }}\,{\ _{t}}=\mathit{\ 20438\ words.\ 
Exceeds\ grOptionDisplayLimit}
\]
\end{maplelatex}

\begin{maplelatex}
\[
{\mathit{R\ }_{\phi }}\,{\ _{\phi }}=\mathit{\ 72053\ words.\ 
Exceeds\ grOptionDisplayLimit}
\]
\end{maplelatex}

\begin{maplelatex}
\[
{\mathit{R\ }_{t}}\,{\ _{t}}=\mathit{\ 82672\ words.\ Exceeds\ 
grOptionDisplayLimit}
\]
\end{maplelatex}

\end{maplegroup}
\begin{maplegroup}

\bigskip
For completeness, the definitions of the Legendre function,
$P_2(\cos(\theta))$, that we will need to calculate the
projections listed in Table 1, and the standard Zerilli substitutions,
that we will need to eliminate $H_1$, and, $\dot K$,
appear below.

\bigskip

\begin{mapleinput}
\mapleinline{active}{1d}{p2 := 5/2*(3*cos(theta)^2-1)/2;}{%
}
\end{mapleinput}

\mapleresult
\begin{maplelatex}
\[
\mathit{p2} := {\displaystyle \frac {15}{4}} \,\mathrm{cos}(
\theta )^{2} - {\displaystyle \frac {5}{4}} 
\]
\end{maplelatex}

\end{maplegroup}
\begin{maplegroup}
\begin{mapleinput}
\mapleinline{active}{1d}{newH(r,t) := }{%
}
\end{mapleinput}

\begin{mapleinput}

\mapleinline{active}{1d}{(2*r^2-6*m*r-3*m^2)/(r-2*m)/(2*r+3*m)*chi(r,t)+r^2/(r-2*m)*eta(r,t);
}{%
}
\end{mapleinput}

\mapleresult
\begin{maplelatex}
\[
\mathrm{newH}(r, \,t) := {\displaystyle \frac {(2\,r^{2} - 6\,m\,
r - 3\,m^{2})\,\chi (r, \,t)}{(r - 2\,m)\,(2\,r + 3\,m)}}  + 
{\displaystyle \frac {r^{2}\,\eta (r, \,t)}{r - 2\,m}} 
\]
\end{maplelatex}

\end{maplegroup}
\begin{maplegroup}
\begin{mapleinput}
\mapleinline{active}{1d}{Kdot(r,t) :=
6*(r^2+m*r+m^2)/r^2/(2*r+3*m)*chi(r,t)+eta(r,t);}{%
}
\end{mapleinput}

\mapleresult
\begin{maplelatex}
\[
\mathrm{Kdot}(r, \,t) := 6\,{\displaystyle \frac {(r^{2} + m\,r
 + m^{2})\,\chi (r, \,t)}{r^{2}\,(2\,r + 3\,m)}}  + \eta (r, \,t)
\]
\end{maplelatex}

\end{maplegroup}
\begin{maplegroup}

Following the prescription set out in Table 1, we can now extract the
linear contributions to the seven non-trivial Ricci components. The resulting
PDEs are called $leqn_{1..7}$.

\bigskip

\begin{mapleinput}
\mapleinline{active}{1d}{a1:=coeff(collect(               }{%
}
\end{mapleinput}

\begin{mapleinput}

\mapleinline{active}{1d}{grcomponent(R(dn,dn),[theta,theta]),epsilon),epsilon,1):
}{%
}
\end{mapleinput}

\end{maplegroup}
\begin{maplegroup}
\begin{mapleinput}

\mapleinline{active}{1d}{g1:=coeff(collect(grcomponent(R(dn,dn),[phi,phi]),epsilon),epsilon,1):
}{%
}
\end{mapleinput}

\end{maplegroup}
\begin{maplegroup}
\begin{mapleinput}
\mapleinline{active}{1d}{ag1:=int(expand(  }{%
}
\end{mapleinput}

\begin{mapleinput}

\mapleinline{active}{1d}{a1-g1/sin(theta)^2)*(2*cot(theta)*diff(p2,theta)+2*(2+1)*p2)*sin(theta)
   }{%
}
\end{mapleinput}

\begin{mapleinput}
\mapleinline{active}{1d}{,theta=0..Pi);}{%
}
\end{mapleinput}

\mapleresult
\begin{maplelatex}
\[
\mathit{ag1} := 12\,{h_{2}}(r, \,t) - 12\,{h_{0}}(r, \,t)
\]
\end{maplelatex}

\end{maplegroup}
\begin{maplegroup}
\begin{mapleinput}
\mapleinline{active}{1d}{leqn[1]:= \{h[2](r,t) =
solve(ag1,h[2](r,t))\};}{%
}
\end{mapleinput}

\mapleresult
\begin{maplelatex}
\[
{\mathit{leqn}_{1}} := \{{h_{2}}(r, \,t)={h_{0}}(r, \,t)\}
\]
\end{maplelatex}

\end{maplegroup}
\begin{maplegroup}
\begin{mapleinput}
\mapleinline{active}{1d}{a2:=subs(leqn[1],a1):}{%
}
\end{mapleinput}

\end{maplegroup}
\begin{maplegroup}
\begin{mapleinput}
\mapleinline{active}{1d}{g2:=subs(leqn[1],g1):}{%
}
\end{mapleinput}

\end{maplegroup}
\begin{maplegroup}
\begin{mapleinput}
\mapleinline{active}{1d}{grmap(R(dn,dn),subs,leqn[1],'x');}{%
}
\end{mapleinput}

\mapleresult
\begin{maplettyout}
Applying routine subs to R(dn,dn)
\end{maplettyout}

\end{maplegroup}
\begin{maplegroup}
\begin{mapleinput}
\mapleinline{active}{1d}{gralter(R(dn,dn),simplify,expand);}{%
}
\end{mapleinput}

\mapleresult
\begin{maplettyout}
Component simplification of a GRTensorII object:
\end{maplettyout}

\begin{maplettyout}
 
\end{maplettyout}

\begin{maplettyout}
Applying routine simplify to object R(dn,dn)
\end{maplettyout}

\begin{maplettyout}
Applying routine expand to object R(dn,dn)
\end{maplettyout}

\begin{maplelatex}
\[
\mathit{CPU\ Time\ }=14.021
\]
\end{maplelatex}

\end{maplegroup}
\begin{maplegroup}
\begin{mapleinput}

\mapleinline{active}{1d}{b1:=coeff(collect(grcomponent(R(dn,dn),[theta,r]),epsilon),epsilon,1):
}{%
}
\end{mapleinput}

\end{maplegroup}
\begin{maplegroup}
\begin{mapleinput}
\mapleinline{active}{1d}{leqn[2]:=termsimp(    }{%
}
\end{mapleinput}

\begin{mapleinput}

\mapleinline{active}{1d}{collect(1/3*int(b1*diff(p2,theta)*sin(theta),theta=0..Pi),lperts));
}{%
}
\end{mapleinput}

\mapleresult
\begin{maplelatex}
\[
{\mathit{leqn}_{2}} :=  - 2\,{\displaystyle \frac {m\,{h_{0}}(r, 
\,t)}{( - r + 2\,m)\,r}}  + ({\frac {\partial }{\partial r}}\,{h
_{0}}(r, \,t)) - ({\frac {\partial }{\partial r}}\,\mathrm{k}(r, 
\,t)) + {\displaystyle \frac {r\,({\frac {\partial }{\partial t}}
\,{h_{1}}(r, \,t))}{ - r + 2\,m}} 
\]
\end{maplelatex}

\end{maplegroup}
\begin{maplegroup}
\begin{mapleinput}

\mapleinline{active}{1d}{c1:=coeff(collect(grcomponent(R(dn,dn),[theta,t]),epsilon),epsilon,1):
}{%
}
\end{mapleinput}

\end{maplegroup}
\begin{maplegroup}
\begin{mapleinput}
\mapleinline{active}{1d}{leqn[3]:=termsimp(    }{%
}
\end{mapleinput}

\begin{mapleinput}

\mapleinline{active}{1d}{collect(1/3*int(c1*diff(p2,theta)*sin(theta),theta=0..Pi),lperts));
}{%
}
\end{mapleinput}

\mapleresult
\begin{maplelatex}
\[
{\mathit{leqn}_{3}} := 2\,{\displaystyle \frac {{h_{1}}(r, \,t)\,
m}{r^{2}}}  - {\displaystyle \frac {( - r + 2\,m)\,({\frac {
\partial }{\partial r}}\,{h_{1}}(r, \,t))}{r}}  - ({\frac {
\partial }{\partial t}}\,{h_{0}}(r, \,t)) - ({\frac {\partial }{
\partial t}}\,\mathrm{k}(r, \,t))
\]
\end{maplelatex}

\end{maplegroup}
\begin{maplegroup}
\begin{mapleinput}

\mapleinline{active}{1d}{d1:=coeff(collect(grcomponent(R(dn,dn),[t,r]),epsilon),epsilon,1):
}{%
}
\end{mapleinput}

\end{maplegroup}
\begin{maplegroup}
\begin{mapleinput}

\mapleinline{active}{1d}{leqn[4]:=termsimp(collect(int(d1*p2*sin(theta),theta=0..Pi),lperts));
}{%
}
\end{mapleinput}

\mapleresult
\begin{maplelatex}
\[
{\mathit{leqn}_{4}} := 3\,{\displaystyle \frac {{h_{1}}(r, \,t)}{
r^{2}}}  - ({\frac {\partial ^{2}}{\partial t\,\partial r}}\,
\mathrm{k}(r, \,t)) + {\displaystyle \frac {{\frac {\partial }{
\partial t}}\,{h_{0}}(r, \,t)}{r}}  - {\displaystyle \frac {( - r
 + 3\,m)\,({\frac {\partial }{\partial t}}\,\mathrm{k}(r, \,t))}{
r\,( - r + 2\,m)}} 
\]
\end{maplelatex}

\end{maplegroup}
\begin{maplegroup}
\begin{mapleinput}

\mapleinline{active}{1d}{e1:=coeff(collect(grcomponent(R(dn,dn),[r,r]),epsilon),epsilon,1):
}{%
}
\end{mapleinput}

\end{maplegroup}
\begin{maplegroup}
\begin{mapleinput}

\mapleinline{active}{1d}{leqn[5]:=termsimp(collect(int(e1*p2*sin(theta),theta=0..Pi),lperts));
}{%
}
\end{mapleinput}

\mapleresult
\begin{maplelatex}
\begin{eqnarray*}
\lefteqn{{\mathit{leqn}_{5}} :=  - 3\,{\displaystyle \frac {{h_{0
}}(r, \,t)}{( - r + 2\,m)\,r}}  + {\displaystyle \frac {1}{2}} \,
({\frac {\partial ^{2}}{\partial r^{2}}}\,{h_{0}}(r, \,t)) - (
{\frac {\partial ^{2}}{\partial r^{2}}}\,\mathrm{k}(r, \,t)) + 
{\displaystyle \frac {r\,({\frac {\partial ^{2}}{\partial t\,
\partial r}}\,{h_{1}}(r, \,t))}{ - r + 2\,m}}  + {\displaystyle 
\frac {1}{2}} \,{\displaystyle \frac {r^{2}\,({\frac {\partial ^{
2}}{\partial t^{2}}}\,{h_{0}}(r, \,t))}{( - r + 2\,m)^{2}}} } \\
 & & \mbox{} - {\displaystyle \frac {{\frac {\partial }{\partial 
r}}\,{h_{0}}(r, \,t)}{ - r + 2\,m}}  - {\displaystyle \frac {(
{\frac {\partial }{\partial r}}\,\mathrm{k}(r, \,t))\,(3\,m - 2\,
r)}{( - r + 2\,m)\,r}}  - {\displaystyle \frac {m\,({\frac {
\partial }{\partial t}}\,{h_{1}}(r, \,t))}{( - r + 2\,m)^{2}}} 
\mbox{\hspace{146pt}}
\end{eqnarray*}
\end{maplelatex}

\end{maplegroup}
\begin{maplegroup}
\begin{mapleinput}

\mapleinline{active}{1d}{f1:=coeff(collect(grcomponent(R(dn,dn),[t,t]),epsilon),epsilon,1):
}{%
}
\end{mapleinput}

\end{maplegroup}
\begin{maplegroup}
\begin{mapleinput}

\mapleinline{active}{1d}{leqn[6]:=termsimp(collect(int(f1*p2*sin(theta),theta=0..Pi),lperts));
}{%
}
\end{mapleinput}

\mapleresult
\begin{maplelatex}
\begin{eqnarray*}
\lefteqn{{\mathit{leqn}_{6}} :=  - 3\,{\displaystyle \frac {( - r
 + 2\,m)\,{h_{0}}(r, \,t)}{r^{3}}}  - {\displaystyle \frac {1}{2}
} \,{\displaystyle \frac {( - r + 2\,m)^{2}\,({\frac {\partial ^{
2}}{\partial r^{2}}}\,{h_{0}}(r, \,t))}{r^{2}}}  - ({\frac {
\partial ^{2}}{\partial t^{2}}}\,\mathrm{k}(r, \,t))} \\
 & & \mbox{} - {\displaystyle \frac {( - r + 2\,m)\,({\frac {
\partial ^{2}}{\partial t\,\partial r}}\,{h_{1}}(r, \,t))}{r}} 
 - {\displaystyle \frac {1}{2}} \,({\frac {\partial ^{2}}{
\partial t^{2}}}\,{h_{0}}(r, \,t)) + {\displaystyle \frac {( - r
 + 2\,m)\,({\frac {\partial }{\partial r}}\,{h_{0}}(r, \,t))}{r^{
2}}}  \\
 & & \mbox{} - {\displaystyle \frac {m\,( - r + 2\,m)\,({\frac {
\partial }{\partial r}}\,\mathrm{k}(r, \,t))}{r^{3}}}  - 
{\displaystyle \frac {(3\,m - 2\,r)\,({\frac {\partial }{
\partial t}}\,{h_{1}}(r, \,t))}{r^{2}}} 
\end{eqnarray*}
\end{maplelatex}

\end{maplegroup}
\begin{maplegroup}
\begin{mapleinput}
\mapleinline{active}{1d}{leqn[7]:=termsimp(collect(  }{%
}
\end{mapleinput}

\begin{mapleinput}

\mapleinline{active}{1d}{int((a2+g2/sin(theta)^2)*p2*sin(theta),theta=0..Pi),lperts));
}{%
}
\end{mapleinput}

\mapleresult
\begin{maplelatex}
\begin{eqnarray*}
\lefteqn{{\mathit{leqn}_{7}} := 4\,\mathrm{k}(r, \,t) + 2\,{h_{0}
}(r, \,t) + ({\frac {\partial ^{2}}{\partial r^{2}}}\,\mathrm{k}(
r, \,t))\,( - r + 2\,m)\,r - {\displaystyle \frac {r^{3}\,(
{\frac {\partial ^{2}}{\partial t^{2}}}\,\mathrm{k}(r, \,t))}{ - 
r + 2\,m}}  - 2\,({\frac {\partial }{\partial r}}\,{h_{0}}(r, \,t
))\,( - r + 2\,m)} \\
 & & \mbox{} + 2\,({\frac {\partial }{\partial r}}\,\mathrm{k}(r
, \,t))\,(3\,m - 2\,r) - 2\,({\frac {\partial }{\partial t}}\,{h
_{1}}(r, \,t))\,r\mbox{\hspace{221pt}}
\end{eqnarray*}
\end{maplelatex}

\end{maplegroup}
\begin{maplegroup}

\bigskip
\noindent While this completes our formal analysis of the linear
problem, these seven equations, $leqn_{1..7}$, must
now be combined in some unobvious ways in order to facilitate the
reduction of the source terms in the second order calculations. These details
can be found in the ``complete'' worksheet at the GRTensor website.

\bigskip
In order to determine the second order PDEs
($qeqn_{1..7}$) we simply repeat the linear analysis,
but now selecting only the $\epsilon^2$ components of the
Ricci tensor. The second order perturbation functions will,
necessarily, appear in these PDEs in exactly the same manner as the
linear perturbation functions appear in the $leqn_i$
equations. 

\bigskip

\end{maplegroup}
\begin{maplegroup}
\begin{mapleinput}
\mapleinline{active}{1d}{aa1:=simplify(coeff(collect(    }{%
}
\end{mapleinput}

\begin{mapleinput}

\mapleinline{active}{1d}{grcomponent(R(dn,dn),[theta,theta]),epsilon),epsilon,2),trig):
}{%
}
\end{mapleinput}

\end{maplegroup}
\begin{maplegroup}
\begin{mapleinput}

\mapleinline{active}{1d}{gg1:=simplify(expand(1/sin(theta)^2*(coeff(collect(
    }{%
}
\end{mapleinput}

\begin{mapleinput}

\mapleinline{active}{1d}{grcomponent(R(dn,dn),[phi,phi]),epsilon),epsilon,2))),trig):
}{%
}
\end{mapleinput}

\end{maplegroup}
\begin{maplegroup}
\begin{mapleinput}
\mapleinline{active}{1d}{zz1:=simplify(expand(gg1-aa1));}{%
}
\end{mapleinput}

\mapleresult
\begin{maplelatex}
\begin{eqnarray*}
\lefteqn{\mathit{zz1} := {\displaystyle \frac {3}{2}} \,{H_{2}}(r
, \,t) - {\displaystyle \frac {3}{2}} \,{H_{0}}(r, \,t) - 
{\displaystyle \frac {3}{2}} \,{h_{1}}(r, \,t)^{2} + 
{\displaystyle \frac {3}{2}} \,{h_{0}}(r, \,t)^{2} + 
{\displaystyle \frac {21}{2}} \,{h_{1}}(r, \,t)^{2}\,\mathrm{cos}
(\theta )^{2} - {\displaystyle \frac {21}{2}} \,\mathrm{cos}(
\theta )^{2}\,{h_{0}}(r, \,t)^{2}} \\
 & & \mbox{} + 9\,\mathrm{cos}(\theta )^{4}\,{h_{0}}(r, \,t)^{2}
 + {\displaystyle \frac {3}{2}} \,{H_{0}}(r, \,t)\,\mathrm{cos}(
\theta )^{2} - {\displaystyle \frac {3}{2}} \,{H_{2}}(r, \,t)\,
\mathrm{cos}(\theta )^{2} - 9\,{h_{1}}(r, \,t)^{2}\,\mathrm{cos}(
\theta )^{4}\mbox{\hspace{35pt}}
\end{eqnarray*}
\end{maplelatex}

\end{maplegroup}
\begin{maplegroup}
\begin{mapleinput}

\mapleinline{active}{1d}{zz2:=int(zz1*(2*cot(theta)*diff(p2,theta)+2*(2+1)*p2)*sin(theta),
          }{%
}
\end{mapleinput}

\begin{mapleinput}
\mapleinline{active}{1d}{theta=0..Pi);}{%
}
\end{mapleinput}

\mapleresult
\begin{maplelatex}
\[
\mathit{zz2} :=  - 12\,{H_{2}}(r, \,t) - {\displaystyle \frac {12
}{7}} \,{h_{0}}(r, \,t)^{2} + {\displaystyle \frac {12}{7}} \,{h
_{1}}(r, \,t)^{2} + 12\,{H_{0}}(r, \,t)
\]
\end{maplelatex}

\end{maplegroup}
\begin{maplegroup}
\begin{mapleinput}
\mapleinline{active}{1d}{qeqn1:= \{H[2](r,t) =
solve(zz2,H[2](r,t))\};}{%
}
\end{mapleinput}

\mapleresult
\begin{maplelatex}
\[
\mathit{qeqn1} := \{{H_{2}}(r, \,t)= - {\displaystyle \frac {1}{7
}} \,{h_{0}}(r, \,t)^{2} + {\displaystyle \frac {1}{7}} \,{h_{1}}
(r, \,t)^{2} + {H_{0}}(r, \,t)\}
\]
\end{maplelatex}

\end{maplegroup}
\begin{maplegroup}
\begin{mapleinput}
\mapleinline{active}{1d}{qeqn[1]:=qeqn1;}{%
}
\end{mapleinput}

\mapleresult
\begin{maplelatex}
\[
{\mathit{qeqn}_{1}} := \{{H_{2}}(r, \,t)= - {\displaystyle 
\frac {1}{7}} \,{h_{0}}(r, \,t)^{2} + {\displaystyle \frac {1}{7}
} \,{h_{1}}(r, \,t)^{2} + {H_{0}}(r, \,t)\}
\]
\end{maplelatex}

\end{maplegroup}
\begin{maplegroup}
\begin{mapleinput}
\mapleinline{active}{1d}{grmap(R(dn,dn),subs,qeqn1,`x`); }{%
}
\end{mapleinput}

\mapleresult
\begin{maplettyout}
Applying routine subs to R(dn,dn)
\end{maplettyout}

\end{maplegroup}
\begin{maplegroup}
\begin{mapleinput}
\mapleinline{active}{1d}{aa1:=simplify(coeff(collect(    }{%
}
\end{mapleinput}

\begin{mapleinput}

\mapleinline{active}{1d}{grcomponent(R(dn,dn),[theta,theta]),epsilon),epsilon,2),trig):
}{%
}
\end{mapleinput}

\end{maplegroup}
\begin{maplegroup}
\begin{mapleinput}

\mapleinline{active}{1d}{gg1:=simplify(expand(1/sin(theta)^2*(coeff(collect(
    }{%
}
\end{mapleinput}

\begin{mapleinput}

\mapleinline{active}{1d}{grcomponent(R(dn,dn),[phi,phi]),epsilon),epsilon,2))),trig):
}{%
}
\end{mapleinput}

\end{maplegroup}
\begin{maplegroup}
\begin{mapleinput}
\mapleinline{active}{1d}{qeqn[2]:=termsimp(collect(             }{%
}
\end{mapleinput}

\begin{mapleinput}

\mapleinline{active}{1d}{int((gg1+aa1)*p2*sin(theta),theta=0..Pi),qperts)):
}{%
}
\end{mapleinput}

\end{maplegroup}
\begin{maplegroup}
\begin{mapleinput}
\mapleinline{active}{1d}{bb1:=simplify(coeff(collect(   }{%
}
\end{mapleinput}

\begin{mapleinput}

\mapleinline{active}{1d}{grcomponent(R(dn,dn),[theta,r]),epsilon),epsilon,2),trig):
}{%
}
\end{mapleinput}

\end{maplegroup}
\begin{maplegroup}
\begin{mapleinput}
\mapleinline{active}{1d}{qeqn[3]:=termsimp(collect(       }{%
}
\end{mapleinput}

\begin{mapleinput}

\mapleinline{active}{1d}{int(bb1*diff(p2,theta)*sin(theta),theta=0..Pi),qperts)):
}{%
}
\end{mapleinput}

\end{maplegroup}
\begin{maplegroup}
\begin{mapleinput}
\mapleinline{active}{1d}{cc1:=simplify(coeff(collect(   }{%
}
\end{mapleinput}

\begin{mapleinput}

\mapleinline{active}{1d}{grcomponent(R(dn,dn),[t,theta]),epsilon),epsilon,2),trig):
}{%
}
\end{mapleinput}

\end{maplegroup}
\begin{maplegroup}
\begin{mapleinput}
\mapleinline{active}{1d}{qeqn[4]:=termsimp(collect(   }{%
}
\end{mapleinput}

\begin{mapleinput}

\mapleinline{active}{1d}{1/3*int(cc1*diff(p2,theta)*sin(theta),theta=0..Pi),qperts)):
}{%
}
\end{mapleinput}

\end{maplegroup}
\begin{maplegroup}
\begin{mapleinput}
\mapleinline{active}{1d}{dd1:=simplify(coeff(collect(       }{%
}
\end{mapleinput}

\begin{mapleinput}

\mapleinline{active}{1d}{grcomponent(R(dn,dn),[r,t]),epsilon),epsilon,2),trig):
}{%
}
\end{mapleinput}

\end{maplegroup}
\begin{maplegroup}
\begin{mapleinput}

\mapleinline{active}{1d}{qeqn[5]:=termsimp(collect(int(dd1*p2*sin(theta),theta=0..Pi),qperts)):
}{%
}
\end{mapleinput}

\end{maplegroup}
\begin{maplegroup}
\begin{mapleinput}
\mapleinline{active}{1d}{ee1:=simplify(coeff(collect(       }{%
}
\end{mapleinput}

\begin{mapleinput}

\mapleinline{active}{1d}{grcomponent(R(dn,dn),[r,r]),epsilon),epsilon,2),trig):
}{%
}
\end{mapleinput}

\end{maplegroup}
\begin{maplegroup}
\begin{mapleinput}
\mapleinline{active}{1d}{qeqn[6]:=termsimp(collect(           }{%
}
\end{mapleinput}

\begin{mapleinput}

\mapleinline{active}{1d}{expand(int(ee1*p2*sin(theta),theta=0..Pi)),qperts)):
}{%
}
\end{mapleinput}

\end{maplegroup}
\begin{maplegroup}
\begin{mapleinput}
\mapleinline{active}{1d}{qsub1:=
\{diff(H[0](r,t),t)=solve(qeqn[4],diff(H[0](r,t),t))\};}{%
}
\end{mapleinput}

\mapleresult
\begin{maplelatex}
\begin{eqnarray*}
\lefteqn{\mathit{qsub1} := \{{\frac {\partial }{\partial t}}\,{H
_{0}}(r, \,t)={\displaystyle \frac {1}{7}} (14\,{H_{1}}(r, \,t)\,
m + 7\,({\frac {\partial }{\partial r}}\,{H_{1}}(r, \,t))\,r^{2}
 - 14\,({\frac {\partial }{\partial r}}\,{H_{1}}(r, \,t))\,m\,r
 - 7\,({\frac {\partial }{\partial t}}\,\mathrm{K}(r, \,t))\,r^{2
}} \\
 & & \mbox{} + 3\,{h_{0}}(r, \,t)\,({\frac {\partial }{\partial t
}}\,{h_{0}}(r, \,t))\,r^{2} - 3\,{h_{1}}(r, \,t)\,({\frac {
\partial }{\partial t}}\,{h_{1}}(r, \,t))\,r^{2} + 2\,\mathrm{k}(
r, \,t)\,({\frac {\partial }{\partial t}}\,\mathrm{k}(r, \,t))\,r
^{2})/r^{2}\}\mbox{\hspace{49pt}}
\end{eqnarray*}
\end{maplelatex}

\end{maplegroup}
\begin{maplegroup}

\begin{mapleinput}
\mapleinline{active}{1d}{qeqn[1];}{%
}
\end{mapleinput}

\mapleresult
\begin{maplelatex}
\[
\{{H_{2}}(r, \,t)= - {\displaystyle \frac {1}{7}} \,{h_{0}}(r, \,
t)^{2} + {\displaystyle \frac {1}{7}} \,{h_{1}}(r, \,t)^{2} + {H
_{0}}(r, \,t)\}
\]
\end{maplelatex}

\end{maplegroup}
\begin{maplegroup}
\begin{mapleinput}
\mapleinline{active}{1d}{undiff(qeqn[4]);}{%
}
\end{mapleinput}

\mapleresult
\begin{maplelatex}
\begin{eqnarray*}
\lefteqn{2\,{\displaystyle \frac {{H_{1}}(r, \,t)\,m}{r^{2}}}  - 
{\displaystyle \frac {( - r + 2\,m)\,({\frac {\partial }{
\partial r}}\,{H_{1}}(r, \,t))}{r}}  - ({\frac {\partial }{
\partial t}}\,{H_{0}}(r, \,t)) - ({\frac {\partial }{\partial t}}
\,\mathrm{K}(r, \,t)) + {\displaystyle \frac {3}{14}} \,({\frac {
\partial }{\partial t}}\,{h_{0}}(r, \,t)^{2})} \\
 & & \mbox{} - {\displaystyle \frac {3}{14}} \,({\frac {\partial 
}{\partial t}}\,{h_{1}}(r, \,t)^{2}) + {\displaystyle \frac {1}{7
}} \,({\frac {\partial }{\partial t}}\,\mathrm{k}(r, \,t)^{2})
\mbox{\hspace{203pt}}
\end{eqnarray*}
\end{maplelatex}

\end{maplegroup}
\begin{maplegroup}

\bigskip
\noindent The derivation of the first Zerilli equation begins with
the elimination of $\dot H_0$, using $qeqn_4$, in
the $R^{(2)}_{rt}$ component of the Ricci
tensor. We then make the previously defined Zerilli substitutions, and
make various substitutions involving the $leqn_i$
equations\footnote{The actual substitutions are somewhat
arbitrary. We choose to eliminate certain perturbation functions so as
to reproduce the results presented in [1].}. The
resulting expression is called $etaeqn$ in our Maple worksheet. To
simplify this rather awkward expression, we employ a new routine,
{\bf mcollect()}, that collects specified terms that are
multiplied together. This routine could, for example, collect together
terms that have a factor of, $ab$, but not, say, $a^2$, or
$b^2$. A more complete description of this routine appears in
Appendix A. (Readers who are familiar with the computer algebra
environment will, no doubt, already see the utility of such a
procedure.)

\bigskip

\begin{mapleinput}
\mapleinline{active}{1d}{eta1:=qeqn[5]:}{%
}
\end{mapleinput}

\end{maplegroup}
\begin{maplegroup}
\begin{mapleinput}

\mapleinline{active}{1d}{eta2:=termsimp(collect(subs(qsub1,eta1),qperts)):
}{%
}
\end{mapleinput}

\end{maplegroup}
\begin{maplegroup}
\begin{mapleinput}
\mapleinline{active}{1d}{eta3:=termsimp(collect(simplify(       }{%
}
\end{mapleinput}

\begin{mapleinput}

\mapleinline{active}{1d}{subs(H[1](r,t)=newH(r,t),diff(K(r,t),t)=Kdot(r,t),
}{%
}
\end{mapleinput}

\begin{mapleinput}

\mapleinline{active}{1d}{diff(K(r,t),r,t)=diff(Kdot(r,t),r),eta2)),zfuncs)):
}{%
}
\end{mapleinput}

\end{maplegroup}
\begin{maplegroup}
\begin{mapleinput}

\mapleinline{active}{1d}{eta4:=eta(r,t)-termsimp(collect(solve(eta3,eta(r,t)),zfuncs)):
}{%
}
\end{mapleinput}

\end{maplegroup}
\begin{maplegroup}
\begin{mapleinput}
\mapleinline{active}{1d}{eta5:=termsimp(collect( }{%
}
\end{mapleinput}

\begin{mapleinput}
\mapleinline{active}{1d}{subs(
\{diff(h[0](r,t),r,r)=solve(leqn[5],diff(h[0](r,t),r,r))\},eta4),}{%
}
\end{mapleinput}

\begin{mapleinput}
\mapleinline{active}{1d}{zfuncs)):}{%
}
\end{mapleinput}

\end{maplegroup}
\begin{maplegroup}
\begin{mapleinput}
\mapleinline{active}{1d}{eta6:=termsimp(collect(        }{%
}
\end{mapleinput}

\begin{mapleinput}
\mapleinline{active}{1d}{subs(
\{diff(h[0](r,t),r)=solve(leqn[2],diff(h[0](r,t),r)),}{%
}
\end{mapleinput}

\begin{mapleinput}

\mapleinline{active}{1d}{diff(k(r,t),r,t)=solve(leqn[4],diff(k(r,t),r,t))
\},eta5),zfuncs)):}{%
}
\end{mapleinput}

\end{maplegroup}
\begin{maplegroup}
\begin{mapleinput}
\mapleinline{active}{1d}{ug:=
\{diff(k(r,t),r,r)=termsimp(collect(solve(                    }{%
}
\end{mapleinput}

\begin{mapleinput}

\mapleinline{active}{1d}{subs(diff(h[0](r,t),r)=solve(leqn[2],diff(h[0](r,t),r)),
}{%
}
\end{mapleinput}

\begin{mapleinput}

\mapleinline{active}{1d}{subs(lsub1,solve(leqn[6],diff(h[0](r,t),r,r))-
}{%
}
\end{mapleinput}

\begin{mapleinput}

\mapleinline{active}{1d}{solve(leqn[5],diff(h[0](r,t),r,r)))),diff(k(r,t),r,r)),lperts))
\}:}{%
}
\end{mapleinput}

\end{maplegroup}
\begin{maplegroup}
\begin{mapleinput}

\mapleinline{active}{1d}{eta7:=termsimp(collect(subs(ug,eta6),zfuncs)):
}{%
}
\end{mapleinput}

\end{maplegroup}
\begin{maplegroup}
\begin{mapleinput}
\mapleinline{active}{1d}{etaeqn:=eta7:}{%
}
\end{mapleinput}

\end{maplegroup}
\begin{maplegroup}
\bigskip

In order to produce a simplified final expression, we extract the
source terms from $etaeqn$ (by setting the second order
perturbation functions to zero) and then simplify the result by
repeated application of {\bf mcollect()}. 

\bigskip

\end{maplegroup}
\begin{maplegroup}
\begin{mapleinput}
\mapleinline{active}{1d}{esource1:=mcollect(expand(                   
 }{%
}
\end{mapleinput}

\begin{mapleinput}
\mapleinline{active}{1d}{eval(subs(eta(r,t)=0,chi(r,t)=0,etaeqn))),
\{h[1](r,t),h[0](r,t)\}):}{%
}
\end{mapleinput}

\end{maplegroup}
\begin{maplegroup}
\begin{mapleinput}
\mapleinline{active}{1d}{esource2:=mcollect(esource1,
\{h[1](r,t),diff(h[1](r,t),t)\}):}{%
}
\end{mapleinput}

\end{maplegroup}
\begin{maplegroup}
\begin{mapleinput}
\mapleinline{active}{1d}{esource3:=mcollect(esource2,
\{h[1](r,t),diff(k(r,t),r)\}):}{%
}
\end{mapleinput}

\end{maplegroup}
\begin{maplegroup}
\begin{mapleinput}
\mapleinline{active}{1d}{esource4:=mcollect(esource3,
\{h[0](r,t),diff(k(r,t),t)\}):}{%
}
\end{mapleinput}

\end{maplegroup}
\begin{maplegroup}
\begin{mapleinput}
\mapleinline{active}{1d}{esource5:=mcollect(esource4,
\{diff(h[1](r,t),t),diff(k(r,t),t)\}):}{%
}
\end{mapleinput}

\end{maplegroup}
\begin{maplegroup}
\begin{mapleinput}
\mapleinline{active}{1d}{esource6:=mcollect(esource5,
\{k(r,t),h[1](r,t)\}):}{%
}
\end{mapleinput}

\end{maplegroup}
\begin{maplegroup}
\begin{mapleinput}
\mapleinline{active}{1d}{esource7:=mcollect(esource6,
\{k(r,t),diff(k(r,t),t)\}):}{%
}
\end{mapleinput}

\end{maplegroup}
\begin{maplegroup}
\begin{mapleinput}
\mapleinline{active}{1d}{esource8:=mcollect(esource7,
\{diff(k(r,t),r),diff(h[0](r,t),t)\}):}{%
}
\end{mapleinput}

\end{maplegroup}
\begin{maplegroup}
\begin{mapleinput}
\mapleinline{active}{1d}{esource9:=mcollect(esource8,
\{k(r,t),diff(h[0](r,t),t)\}):}{%
}
\end{mapleinput}

\end{maplegroup}
\begin{maplegroup}
\begin{mapleinput}
\mapleinline{active}{1d}{esource10:=mcollect(esource9,
\{h[0](r,t),diff(h[0](r,t),t)\}):}{%
}
\end{mapleinput}

\end{maplegroup}
\begin{maplegroup}
\begin{mapleinput}

\mapleinline{active}{1d}{etasource:=kfactor(esource10,(r-2*m)/7/(2*r+3*m)):
}{%
}
\end{mapleinput}

\end{maplegroup}
\begin{maplegroup}

\bigskip
\noindent It is now a simple matter to put these results together to
arrive at the first Zerilli equation for the second order calculation:

\bigskip
\begin{mapleinput}
\mapleinline{active}{1d}{etafinal:=eta(r,t) = }{%
}
\end{mapleinput}

\begin{mapleinput}

\mapleinline{active}{1d}{termsimp(collect(solve(eval(subs(h[0](r,t)=0,h[1](r,t)=0,k(r,t)=0,
}{%
}
\end{mapleinput}

\begin{mapleinput}
\mapleinline{active}{1d}{etaeqn)),eta(r,t)),zfuncs))-etasource;}{%
}
\end{mapleinput}

\mapleresult
\begin{maplelatex}
\begin{eqnarray*}
\lefteqn{\mathit{etafinal} := \eta (r, \,t)= - {\displaystyle 
\frac {({\frac {\partial }{\partial r}}\,\chi (r, \,t))\,( - r + 
2\,m)}{r}}  - {\displaystyle \frac {1}{7}}  \left( {\vrule 
height1.23em width0em depth1.23em} \right. \!  \! (r - 2\,m)
 \left( {\vrule height1.23em width0em depth1.23em} \right. \! 
 \!  - {h_{0}}(r, \,t)\,({\frac {\partial }{\partial t}}\,{h_{0}}
(r, \,t))} \\
 & & \mbox{} + 2\,\mathrm{k}(r, \,t)\,({\frac {\partial }{
\partial t}}\,{h_{0}}(r, \,t)) + r\,({\frac {\partial }{\partial 
r}}\,\mathrm{k}(r, \,t))\,({\frac {\partial }{\partial t}}\,{h_{0
}}(r, \,t)) + 2\,\mathrm{k}(r, \,t)\,({\frac {\partial }{
\partial t}}\,\mathrm{k}(r, \,t)) + 4\,{\displaystyle \frac {{h_{
1}}(r, \,t)\,\mathrm{k}(r, \,t)}{r}}  \\
 & & \mbox{} - 2\,{\displaystyle \frac {m\,{h_{1}}(r, \,t)\,(
{\frac {\partial }{\partial r}}\,\mathrm{k}(r, \,t))}{r}}  + {h_{
1}}(r, \,t)\,({\frac {\partial }{\partial t}}\,{h_{1}}(r, \,t))
 - 2\,{\displaystyle \frac {( - r + 2\,m)\,{h_{1}}(r, \,t)\,{h_{0
}}(r, \,t)}{r^{2}}}  \\
 & & \mbox{} - 2\,{\displaystyle \frac {m\,({\frac {\partial }{
\partial t}}\,\mathrm{k}(r, \,t))\,{h_{0}}(r, \,t)}{ - r + 2\,m}
}  + {\displaystyle \frac {r^{2}\,({\frac {\partial }{\partial t
}}\,\mathrm{k}(r, \,t))\,({\frac {\partial }{\partial t}}\,{h_{1}
}(r, \,t))}{ - r + 2\,m}}  \! \! \left. {\vrule 
height1.23em width0em depth1.23em} \right)  \! \! \left. {\vrule 
height1.23em width0em depth1.23em} \right) /(2\,r + 3\,m)
\end{eqnarray*}
\end{maplelatex}

\end{maplegroup}
\begin{maplegroup}

\bigskip
\noindent This agrees with the result found in [1].

The development of  the second order Zerilli ``wave equation'' follows
a similar procedure to that of the {\em etafinal} result.
We first take the time derivative of $qeqn_2$, eliminate
$\dot H_0$ and $\dot H_0^\prime$ using
$qeqn_4$, make the Zerilli substitutions, and substitute for
$\eta$ from our first Zerilli result. We then repeat this
process for $qenq_3$ and add the resulting equations such that
the $\ddot \chi^\prime$ term vanishes. The result of these
operations, which is called $chieqn$ in our worksheet, is
essentially the raw form of the result we are seeking. This expression
is, however, rather large.

\bigskip

\begin{mapleinput}
\mapleinline{active}{1d}{atmp1:=diff(qeqn[2],t):}{%
}
\end{mapleinput}

\end{maplegroup}
\begin{maplegroup}
\begin{mapleinput}
\mapleinline{active}{1d}{atmp2:=termsimp(collect(expand(  }{%
}
\end{mapleinput}

\begin{mapleinput}
\mapleinline{active}{1d}{subs(qsub1,diff(qsub1,r),atmp1)),
\{diff(K(r,t),t,t,t)\})):}{%
}
\end{mapleinput}

\end{maplegroup}
\begin{maplegroup}
\begin{mapleinput}

\mapleinline{active}{1d}{atmp3:=termsimp(collect(expand(eval(subs(H[1](r,t)=newH(r,t),
}{%
}
\end{mapleinput}

\begin{mapleinput}
\mapleinline{active}{1d}{diff(K(r,t),t)=Kdot(r,t),diff(K(r,t),t,}{%
}
\end{mapleinput}

\begin{mapleinput}
\mapleinline{active}{1d}{r)=diff(Kdot(r,t),r),
diff(K(r,t),t,r,r)=diff(Kdot(r,t),r,r),atmp2))),zfuncs)):}{%
}
\end{mapleinput}

\end{maplegroup}
\begin{maplegroup}
\begin{mapleinput}

\mapleinline{active}{1d}{atmp4:=termsimp(collect(expand(eval(subs(etafinal,atmp3))),zfuncs)):
}{%
}
\end{mapleinput}

\end{maplegroup}
\begin{maplegroup}
\begin{mapleinput}

\mapleinline{active}{1d}{coefa:=simplify(coeff(atmp4,diff(chi(r,t),t,r,t),1));
}{%
}
\end{mapleinput}

\mapleresult
\begin{maplelatex}
\[
\mathit{coefa} :=  - r^{2}
\]
\end{maplelatex}

\end{maplegroup}
\begin{maplegroup}
\begin{mapleinput}
\mapleinline{active}{1d}{btmp1:=termsimp(collect(expand(}{%
}
\end{mapleinput}

\begin{mapleinput}
\mapleinline{active}{1d}{subs(qsub1,diff(qsub1,r),diff(qeqn[3],t))),
\{diff(H[1](r,t),t,t),}{%
}
\end{mapleinput}

\begin{mapleinput}
\mapleinline{active}{1d}{diff(K(r,t),t,t,t)\})):}{%
}
\end{mapleinput}

\end{maplegroup}
\begin{maplegroup}
\begin{mapleinput}
\mapleinline{active}{1d}{btmp2:=termsimp(collect(expand(eval(}{%
}
\end{mapleinput}

\begin{mapleinput}

\mapleinline{active}{1d}{subs(H[1](r,t)=newH(r,t),diff(K(r,t),t)=Kdot(r,t),
}{%
}
\end{mapleinput}

\begin{mapleinput}

\mapleinline{active}{1d}{diff(K(r,t),r,t)=diff(Kdot(r,t),r),btmp1))),zfuncs)):
}{%
}
\end{mapleinput}

\end{maplegroup}
\begin{maplegroup}
\begin{mapleinput}

\mapleinline{active}{1d}{btmp3:=termsimp(collect(expand(eval(subs(etafinal,btmp2))),zfuncs)):
}{%
}
\end{mapleinput}

\end{maplegroup}
\begin{maplegroup}
\begin{mapleinput}

\mapleinline{active}{1d}{coefb:=simplify(coeff(btmp3,diff(chi(r,t),r,t,t),1));
}{%
}
\end{mapleinput}

\mapleresult
\begin{maplelatex}
\[
\mathit{coefb} := 3\,{\displaystyle \frac {r^{2}}{ - r + 2\,m}} 
\]
\end{maplelatex}

\end{maplegroup}
\begin{maplegroup}
\begin{mapleinput}

\mapleinline{active}{1d}{ftmp1:=termsimp(collect(expand(atmp4/coefa-btmp3/coefb),
}{%
}
\end{mapleinput}

\begin{mapleinput}
\mapleinline{active}{1d}{
\{chi(r,t),diff(chi(r,t),r),diff(chi(r,t),t),diff(chi(r,t),t,t),}{%
}
\end{mapleinput}

\begin{mapleinput}

\mapleinline{active}{1d}{diff(chi(r,t),r,r),diff(chi(r,t),r,t),diff(chi(r,t),r,t,t),
}{%
}
\end{mapleinput}

\begin{mapleinput}
\mapleinline{active}{1d}{diff(chi(r,t),r,r,r)\})):}{%
}
\end{mapleinput}

\end{maplegroup}
\begin{maplegroup}
\begin{mapleinput}
\mapleinline{active}{1d}{ftmp2:=termsimp(collect(expand(eval(}{%
}
\end{mapleinput}

\begin{mapleinput}

\mapleinline{active}{1d}{subs(k(r,t)=0,h[1](r,t)=0,h[0](r,t)=0,ftmp1))),
\{chi(r,t),}{%
}
\end{mapleinput}

\begin{mapleinput}

\mapleinline{active}{1d}{diff(chi(r,t),r),diff(chi(r,t),r,r),diff(chi(r,t),t,t),
}{%
}
\end{mapleinput}

\begin{mapleinput}
\mapleinline{active}{1d}{diff(chi(r,t),r,r,r)\})):}{%
}
\end{mapleinput}

\end{maplegroup}
\begin{maplegroup}
\begin{mapleinput}
\mapleinline{active}{1d}{chi1:=termsimp(collect(}{%
}
\end{mapleinput}

\begin{mapleinput}

\mapleinline{active}{1d}{diff(chi(r,t),t,t)-solve(ftmp1,diff(chi(r,t),t,t)),zfuncs)):
}{%
}
\end{mapleinput}

\end{maplegroup}
\begin{maplegroup}
\begin{mapleinput}
\mapleinline{active}{1d}{chieqn:=ftmp1:}{%
}
\end{mapleinput}

\end{maplegroup}
\begin{maplegroup}

\begin{mapleinput}
\mapleinline{active}{1d}{nops(chieqn);}{%
}
\end{mapleinput}

\mapleresult
\begin{maplelatex}
\[
421
\]
\end{maplelatex}

\end{maplegroup}
\begin{maplegroup}

\bigskip
\noindent Although, at this point we can at least verify that it
reproduces the correct linear Zerilli equation when the source terms
are set to zero.

\bigskip

\begin{mapleinput}

\mapleinline{active}{1d}{termsimp(collect(expand(eval(subs(k(r,t)=0,h[1](r,t)=0,h[0](r,t)=0,
}{%
}
\end{mapleinput}

\begin{mapleinput}
\mapleinline{active}{1d}{chieqn/coeff(chieqn,diff(chi(r,t),t,t),1)))),
\{chi(r,t),diff(chi(r,t),r),}{%
}
\end{mapleinput}

\begin{mapleinput}

\mapleinline{active}{1d}{diff(chi(r,t),r,r),diff(chi(r,t),t,t),diff(chi(r,t),r,r,r)
\}));}{%
}
\end{mapleinput}

\mapleresult
\begin{maplelatex}
\begin{eqnarray*}
\lefteqn{ - {\displaystyle \frac {({\frac {\partial ^{2}}{
\partial r^{2}}}\,\chi (r, \,t))\,( - r + 2\,m)^{2}}{r^{2}}}  + (
{\frac {\partial ^{2}}{\partial t^{2}}}\,\chi (r, \,t)) - 6\,
{\displaystyle \frac {( - r + 2\,m)\,(3\,m^{3} + 6\,r\,m^{2} + 4
\,r^{2}\,m + 4\,r^{3})\,\chi (r, \,t)}{r^{4}\,(2\,r + 3\,m)^{2}}
} } \\
 & & \mbox{} + 2\,{\displaystyle \frac {({\frac {\partial }{
\partial r}}\,\chi (r, \,t))\,( - r + 2\,m)\,m}{r^{3}}} 
\mbox{\hspace{262pt}}
\end{eqnarray*}
\end{maplelatex}

\end{maplegroup}
\begin{maplegroup}

\noindent This is just the linear Zerilli wave equation with the
quadrapole potential, as required.

In order to make contact with the results presented in
[1], we now work with the ``renormalized'' function $\zeta$.

\bigskip

\begin{mapleinput}
\mapleinline{active}{1d}{zetasub:= \{chi(r,t) =}{%
}
\end{mapleinput}

\begin{mapleinput}

\mapleinline{active}{1d}{zeta(r,t)+2/7*(r^2/(2*r+3*m)*k(r,t)*diff(k(r,t),t)+k(r,t)^2)
\};}{%
}
\end{mapleinput}

\mapleresult
\begin{maplelatex}
\[
\mathit{zetasub} :=  \left\{  \! \chi (r, \,t)=\zeta (r, \,t) + 
{\displaystyle \frac {2}{7}} \,{\displaystyle \frac {r^{2}\,
\mathrm{k}(r, \,t)\,({\frac {\partial }{\partial t}}\,\mathrm{k}(
r, \,t))}{2\,r + 3\,m}}  + {\displaystyle \frac {2}{7}} \,
\mathrm{k}(r, \,t)^{2} \!  \right\} 
\]
\end{maplelatex}

\end{maplegroup}
\begin{maplegroup}
\begin{mapleinput}
\mapleinline{active}{1d}{zeta1:=eval(subs(zetasub,chi1)):}{%
}
\end{mapleinput}

\end{maplegroup}
\begin{maplegroup}
\begin{mapleinput}
\mapleinline{active}{1d}{nops(expand(zeta1));}{%
}
\end{mapleinput}

\mapleresult
\begin{maplelatex}
\[
161
\]
\end{maplelatex}

\end{maplegroup}
\begin{maplegroup}
\bigskip

\noindent This substitution has considerably reduced the number of
terms in the expression. 
\bigskip
We now eliminate $h_0$ in favor of $\psi$ using:

\bigskip

\end{maplegroup}
\begin{maplegroup}
\begin{mapleinput}
\mapleinline{active}{1d}{psisub:=\{h[0](r,t) = }{%
}
\end{mapleinput}

\begin{mapleinput}
\mapleinline{active}{1d}{(psi(r,t) - r/3*k(r,t))*3*(2*r+3*m)/r/(r-2*m)
+r*diff(k(r,t),r)\};}{%
}
\end{mapleinput}

\mapleresult
\begin{maplelatex}
\[
\mathit{psisub} :=  \left\{  \! {h_{0}}(r, \,t)=3\,
{\displaystyle \frac {(\psi (r, \,t) - {\displaystyle \frac {1}{3
}} \,r\,\mathrm{k}(r, \,t))\,(2\,r + 3\,m)}{r\,(r - 2\,m)}}  + r
\,({\frac {\partial }{\partial r}}\,\mathrm{k}(r, \,t)) \! 
 \right\} 
\]
\end{maplelatex}

\end{maplegroup}
\begin{maplegroup}
\bigskip
\noindent After making this substitution, and a fairly extensive
simplification of the source terms using the linear perturbation
equations\footnote{While we have omitted these details here, they
can be found in the full Maple worksheet at the GRTensor
website.}, we arrive at the unsimplified answer $zetaeqn$.
\bigskip

\begin{mapleinput}
\mapleinline{active}{1d}{nops(zetaeqn);}{%
}
\end{mapleinput}

\mapleresult
\begin{maplelatex}
\[
154
\]
\end{maplelatex}

\end{maplegroup}
\begin{maplegroup}

\bigskip
\noindent Here again we make use of the {\bf mcollect()}
routine to simplify our result.

\bigskip

\begin{mapleinput}
\mapleinline{active}{1d}{final1:=mcollect(expand(zetaeqn),
\{psi(r,t),diff(psi(r,t),r,r)\}):}{%
}
\end{mapleinput}

\end{maplegroup}
\begin{maplegroup}
\begin{mapleinput}
\mapleinline{active}{1d}{final2:=mcollect(final1,
\{psi(r,t),diff(psi(r,t),t)\}):}{%
}
\end{mapleinput}

\end{maplegroup}
\begin{maplegroup}
\begin{mapleinput}
\mapleinline{active}{1d}{final3:=mcollect(final2,
\{diff(psi(r,t),t),diff(psi(r,t),r,r,r)\}):}{%
}
\end{mapleinput}

\end{maplegroup}
\begin{maplegroup}
\begin{mapleinput}
\mapleinline{active}{1d}{final4:=mcollect(final3,
\{diff(psi(r,t),r),diff(psi(r,t),t)\}):}{%
}
\end{mapleinput}

\end{maplegroup}
\begin{maplegroup}
\begin{mapleinput}
\mapleinline{active}{1d}{final5:=mcollect(final4,
\{diff(psi(r,t),t),diff(psi(r,t),t,r)\}):}{%
}
\end{mapleinput}

\end{maplegroup}
\begin{maplegroup}
\begin{mapleinput}
\mapleinline{active}{1d}{final6:=mcollect(final5,
\{psi(r,t),diff(psi(r,t),r)\}):}{%
}
\end{mapleinput}

\end{maplegroup}
\begin{maplegroup}
\begin{mapleinput}
\mapleinline{active}{1d}{final7:=mcollect(final6,
\{diff(psi(r,t),r),diff(psi(r,t),t,r,r)\}):}{%
}
\end{mapleinput}

\end{maplegroup}
\begin{maplegroup}
\begin{mapleinput}
\mapleinline{active}{1d}{final8:=mcollect(final7,
\{diff(psi(r,t),t),diff(psi(r,t),r,r)\}):}{%
}
\end{mapleinput}

\end{maplegroup}
\begin{maplegroup}
\begin{mapleinput}
\mapleinline{active}{1d}{final9:=mcollect(final8,
\{psi(r,t),diff(psi(r,t),r,t)\}):}{%
}
\end{mapleinput}

\end{maplegroup}
\begin{maplegroup}
\begin{mapleinput}
\mapleinline{active}{1d}{final10:=mcollect(final9,
\{diff(psi(r,t),r),diff(psi(r,t),r,t)\}):}{%
}
\end{mapleinput}

\end{maplegroup}
\begin{maplegroup}
\begin{mapleinput}
\mapleinline{active}{1d}{final11:=mcollect(final10,
\{psi(r,t),diff(psi(r,t),t,r,r)\}):}{%
}
\end{mapleinput}

\end{maplegroup}
\begin{maplegroup}
\begin{mapleinput}
\mapleinline{active}{1d}{final12:=mcollect(final11,
\{diff(psi(r,t),r),diff(psi(r,t),r,r)\}):}{%
}
\end{mapleinput}

\end{maplegroup}
\begin{maplegroup}
\begin{mapleinput}
\mapleinline{active}{1d}{final13:=mcollect(final12,\{psi(r,t)\}):}{%
}
\end{mapleinput}

\end{maplegroup}
\begin{maplegroup}
\begin{mapleinput}
\mapleinline{active}{1d}{final14:=mcollect(final13,\{diff(psi(r,t),t)
\}):}{%
}
\end{mapleinput}

\end{maplegroup}
\begin{maplegroup}
\begin{mapleinput}
\mapleinline{active}{1d}{final15:=mcollect(final14,\{diff(psi(r,t),r)
\}):}{%
}
\end{mapleinput}

\end{maplegroup}
\begin{maplegroup}
\begin{mapleinput}
\mapleinline{active}{1d}{final16:=mcollect(final15,
\{diff(psi(r,t),r,r)\}):}{%
}
\end{mapleinput}

\end{maplegroup}
\begin{maplegroup}
\begin{mapleinput}
\mapleinline{active}{1d}{final17:=mcollect(final16,
\{diff(psi(r,t),r,t)\}):}{%
}
\end{mapleinput}

\end{maplegroup}
\begin{maplegroup}
\begin{mapleinput}
\mapleinline{active}{1d}{nops(final17);}{%
}
\end{mapleinput}

\mapleresult
\begin{maplelatex}
\[
42
\]
\end{maplelatex}

\end{maplegroup}
\begin{maplegroup}
\noindent We now make one final set of substitutions,
$\mu=(r-2m)$, and, $\lambda=(2r+3m)$, and extract the
source terms from the second order Zerilli wave equation:

\bigskip

\begin{mapleinput}

\mapleinline{active}{1d}{zetasource:=subs((r-2*m)=mu,(2*m-r)=-mu,(2*r+3*m)=lambda,
}{%
}
\end{mapleinput}

\begin{mapleinput}

\mapleinline{active}{1d}{kfactor(eval(subs(zeta(r,t)=0,-final17)),12/7*(r-2*m)^3/(2*r+3*m)));
}{%
}
\end{mapleinput}

\mapleresult
\begin{maplelatex}
\begin{eqnarray*}
\lefteqn{\mathit{zetasource} := {\displaystyle \frac {12}{7}} \mu
 ^{3} \left( {\vrule height1.37em width0em depth1.37em}
 \right. \!  \!  - {\displaystyle \frac {1}{3}} \,{\displaystyle 
\frac {(7\,m - 3\,r)\,({\frac {\partial }{\partial r}}\,\psi (r, 
\,t))\,({\frac {\partial ^{2}}{\partial t\,\partial r}}\,\psi (r
, \,t))}{\mu \,r^{3}}} } \\
 & & \mbox{} + {\displaystyle \frac {(8\,r^{2} + 12\,m\,r + 7\,m
^{2})\,\psi (r, \,t)\,({\frac {\partial ^{2}}{\partial t\,
\partial r}}\,\psi (r, \,t))}{\mu \,r^{4}\,\lambda }}  \\
 & & \mbox{} - 4\,{\displaystyle \frac {(2\,r^{3} + 4\,r^{2}\,m
 + 9\,r\,m^{2} + 6\,m^{3})\,\psi (r, \,t)\,\mathrm{\%1}}{r^{6}\,
\lambda }}  \\
 & & \mbox{} - 12\,{\displaystyle \frac {( - 2\,r^{5} - 9\,r^{4}
\,m - 6\,r^{3}\,m^{2} + 2\,r^{2}\,m^{3} + 15\,r\,m^{4} + 15\,m^{5
})\,\psi (r, \,t)^{2}}{\mu ^{2}\,r^{8}\,\lambda }}  \\
 & & \mbox{} + {\displaystyle \frac {(112\,r^{5} + 480\,r^{4}\,m
 + 692\,r^{3}\,m^{2} + 762\,r^{2}\,m^{3} + 441\,r\,m^{4} + 144\,m
^{5})\,\psi (r, \,t)\,({\frac {\partial }{\partial t}}\,\psi (r, 
\,t))}{\mu ^{2}\,\lambda ^{3}\,r^{5}}}  \\
 & & \mbox{} - 4\,{\displaystyle \frac {(r^{2} + m\,r + m^{2})\,(
{\frac {\partial }{\partial t}}\,\psi (r, \,t))\,({\frac {
\partial ^{2}}{\partial t\,\partial r}}\,\psi (r, \,t))}{\mu ^{2}
\,r^{3}}}  + {\displaystyle \frac {1}{3}} \,{\displaystyle 
\frac {\mu \,\mathrm{\%1}^{2}\,\lambda }{r^{4}}}  \\
 & & \mbox{} + {\displaystyle \frac {1}{3}} \,{\displaystyle 
\frac {(12\,r^{3} + 36\,r^{2}\,m + 59\,r\,m^{2} + 90\,m^{3})\,(
{\frac {\partial }{\partial r}}\,\psi (r, \,t))^{2}}{\mu \,r^{6}}
}  \\
 & & \mbox{} - 2\,{\displaystyle \frac {(32\,r^{5} + 88\,r^{4}\,m
 + 296\,r^{3}\,m^{2} + 510\,r^{2}\,m^{3} + 561\,r\,m^{4} + 270\,m
^{5})\,\psi (r, \,t)\,({\frac {\partial }{\partial r}}\,\psi (r, 
\,t))}{\mu \,r^{7}\,\lambda ^{2}}}  \\
 & & \mbox{} - {\displaystyle \frac {1}{3}} \,{\displaystyle 
\frac {( - 18\,r^{3} + 4\,r^{2}\,m + 33\,r\,m^{2} + 48\,m^{3})\,(
{\frac {\partial }{\partial t}}\,\psi (r, \,t))\,({\frac {
\partial }{\partial r}}\,\psi (r, \,t))}{\mu ^{2}\,r^{4}\,\lambda
 }}  \\
 & & \mbox{} - {\displaystyle \frac {1}{3}} \,{\displaystyle 
\frac {({\frac {\partial }{\partial t}}\,\psi (r, \,t))\,(
{\frac {\partial ^{3}}{\partial r^{3}}}\,\psi (r, \,t))}{r^{2}}} 
 - {\displaystyle \frac {m\,\psi (r, \,t)\,({\frac {\partial ^{3}
}{\partial t\,\partial r^{2}}}\,\psi (r, \,t))}{r^{3}\,\lambda }
}  + {\displaystyle \frac {(m^{2} - 2\,r^{2})\,({\frac {\partial 
}{\partial t}}\,\psi (r, \,t))\,\mathrm{\%1}}{\mu \,r^{3}\,
\lambda }}  \\
 & & \mbox{} + {\displaystyle \frac {4}{3}} \,{\displaystyle 
\frac {(3\,r^{2} + 5\,m\,r + 6\,m^{2})\,({\frac {\partial }{
\partial r}}\,\psi (r, \,t))\,\mathrm{\%1}}{r^{5}}}  + 
{\displaystyle \frac {1}{3}} \,{\displaystyle \frac {({\frac {
\partial }{\partial r}}\,\psi (r, \,t))\,({\frac {\partial ^{3}}{
\partial t\,\partial r^{2}}}\,\psi (r, \,t))}{r^{2}}}  \\
 & & \mbox{} - 12\,{\displaystyle \frac {(r^{2} + m\,r + m^{2})^{
2}\,({\frac {\partial }{\partial t}}\,\psi (r, \,t))^{2}}{r^{4}\,
\mu ^{3}\,\lambda }}  - {\displaystyle \frac {1}{3}} \,
{\displaystyle \frac {({\frac {\partial ^{2}}{\partial t\,
\partial r}}\,\psi (r, \,t))^{2}\,\lambda }{\mu \,r^{2}}}  \! 
\! \left. {\vrule height1.37em width0em depth1.37em} \right) /
\lambda  \\
 & & \mathrm{\%1} := {\frac {\partial ^{2}}{\partial r^{2}}}\,
\psi (r, \,t)
\end{eqnarray*}
\end{maplelatex}

\end{maplegroup}
\bigskip
\noindent This is the effective source term for the second order Zerilli
equation.